\documentclass[aps,prd,preprint,tightenlines,superscriptaddress]{revtex4}
\usepackage{graphicx} 
\usepackage{dcolumn}  

\usepackage{graphicx,multirow}
\usepackage{xspace}

\usepackage{subfigure}

\usepackage{amsmath,amsfonts}

\usepackage{rotating} 
\usepackage{hyperref}
\usepackage{soul}
\usepackage{float}
\usepackage{bm}
\usepackage{notoccite}

\newcommand{\beq}{\begin{equation}}
\newcommand{\eeq}{\end{equation}}
\newcommand{\beqa}{\begin{eqnarray}}
\newcommand{\eeqa}{\end{eqnarray}}

\newcommand{\Bbar}{\,\overline{\!B}{}}
\newcommand{\Dbar}{\,\overline{\!D}{}}
\newcommand{\Kbar}{\,\overline{\!K}{}}
\def\B0bar{\Bbar{}^0}
\def\D0bar{\Dbar{}^0}
\def\K0bar{\Kbar{}^0}

\graphicspath{{ps}}

\begin{document}


\preprint{\vbox{ \hbox{   }
                 \hbox{BELLE-CONF-1612}
}}

\title{ \quad\\[0.5cm]  \boldmath Precise determination of the CKM matrix element $\left| V_{cb}\right|$ with $\bar B^0 \to D^{*\,+} \, \ell^- \, \bar \nu_\ell$ decays with hadronic tagging at Belle  }


\noaffiliation
\affiliation{Aligarh Muslim University, Aligarh 202002}
\affiliation{University of the Basque Country UPV/EHU, 48080 Bilbao}
\affiliation{Beihang University, Beijing 100191}
\affiliation{University of Bonn, 53115 Bonn}
\affiliation{Budker Institute of Nuclear Physics SB RAS, Novosibirsk 630090}
\affiliation{Faculty of Mathematics and Physics, Charles University, 121 16 Prague}
\affiliation{Chiba University, Chiba 263-8522}
\affiliation{Chonnam National University, Kwangju 660-701}
\affiliation{University of Cincinnati, Cincinnati, Ohio 45221}
\affiliation{Deutsches Elektronen--Synchrotron, 22607 Hamburg}
\affiliation{University of Florida, Gainesville, Florida 32611}
\affiliation{Department of Physics, Fu Jen Catholic University, Taipei 24205}
\affiliation{Justus-Liebig-Universit\"at Gie\ss{}en, 35392 Gie\ss{}en}
\affiliation{Gifu University, Gifu 501-1193}
\affiliation{II. Physikalisches Institut, Georg-August-Universit\"at G\"ottingen, 37073 G\"ottingen}
\affiliation{SOKENDAI (The Graduate University for Advanced Studies), Hayama 240-0193}
\affiliation{Gyeongsang National University, Chinju 660-701}
\affiliation{Hanyang University, Seoul 133-791}
\affiliation{University of Hawaii, Honolulu, Hawaii 96822}
\affiliation{High Energy Accelerator Research Organization (KEK), Tsukuba 305-0801}
\affiliation{J-PARC Branch, KEK Theory Center, High Energy Accelerator Research Organization (KEK), Tsukuba 305-0801}
\affiliation{Hiroshima Institute of Technology, Hiroshima 731-5193}
\affiliation{IKERBASQUE, Basque Foundation for Science, 48013 Bilbao}
\affiliation{University of Illinois at Urbana-Champaign, Urbana, Illinois 61801}
\affiliation{Indian Institute of Science Education and Research Mohali, SAS Nagar, 140306}
\affiliation{Indian Institute of Technology Bhubaneswar, Satya Nagar 751007}
\affiliation{Indian Institute of Technology Guwahati, Assam 781039}
\affiliation{Indian Institute of Technology Madras, Chennai 600036}
\affiliation{Indiana University, Bloomington, Indiana 47408}
\affiliation{Institute of High Energy Physics, Chinese Academy of Sciences, Beijing 100049}
\affiliation{Institute of High Energy Physics, Vienna 1050}
\affiliation{Institute for High Energy Physics, Protvino 142281}
\affiliation{Institute of Mathematical Sciences, Chennai 600113}
\affiliation{INFN - Sezione di Torino, 10125 Torino}
\affiliation{Advanced Science Research Center, Japan Atomic Energy Agency, Naka 319-1195}
\affiliation{J. Stefan Institute, 1000 Ljubljana}
\affiliation{Kanagawa University, Yokohama 221-8686}
\affiliation{Institut f\"ur Experimentelle Kernphysik, Karlsruher Institut f\"ur Technologie, 76131 Karlsruhe}
\affiliation{Kavli Institute for the Physics and Mathematics of the Universe (WPI), University of Tokyo, Kashiwa 277-8583}
\affiliation{Kennesaw State University, Kennesaw, Georgia 30144}
\affiliation{King Abdulaziz City for Science and Technology, Riyadh 11442}
\affiliation{Department of Physics, Faculty of Science, King Abdulaziz University, Jeddah 21589}
\affiliation{Korea Institute of Science and Technology Information, Daejeon 305-806}
\affiliation{Korea University, Seoul 136-713}
\affiliation{Kyoto University, Kyoto 606-8502}
\affiliation{Kyungpook National University, Daegu 702-701}
\affiliation{\'Ecole Polytechnique F\'ed\'erale de Lausanne (EPFL), Lausanne 1015}
\affiliation{P.N. Lebedev Physical Institute of the Russian Academy of Sciences, Moscow 119991}
\affiliation{Faculty of Mathematics and Physics, University of Ljubljana, 1000 Ljubljana}
\affiliation{Ludwig Maximilians University, 80539 Munich}
\affiliation{Luther College, Decorah, Iowa 52101}
\affiliation{University of Malaya, 50603 Kuala Lumpur}
\affiliation{University of Maribor, 2000 Maribor}
\affiliation{Max-Planck-Institut f\"ur Physik, 80805 M\"unchen}
\affiliation{School of Physics, University of Melbourne, Victoria 3010}
\affiliation{Middle East Technical University, 06531 Ankara}
\affiliation{University of Miyazaki, Miyazaki 889-2192}
\affiliation{Moscow Physical Engineering Institute, Moscow 115409}
\affiliation{Moscow Institute of Physics and Technology, Moscow Region 141700}
\affiliation{Graduate School of Science, Nagoya University, Nagoya 464-8602}
\affiliation{Kobayashi-Maskawa Institute, Nagoya University, Nagoya 464-8602}
\affiliation{Nara University of Education, Nara 630-8528}
\affiliation{Nara Women's University, Nara 630-8506}
\affiliation{National Central University, Chung-li 32054}
\affiliation{National United University, Miao Li 36003}
\affiliation{Department of Physics, National Taiwan University, Taipei 10617}
\affiliation{H. Niewodniczanski Institute of Nuclear Physics, Krakow 31-342}
\affiliation{Nippon Dental University, Niigata 951-8580}
\affiliation{Niigata University, Niigata 950-2181}
\affiliation{University of Nova Gorica, 5000 Nova Gorica}
\affiliation{Novosibirsk State University, Novosibirsk 630090}
\affiliation{Osaka City University, Osaka 558-8585}
\affiliation{Osaka University, Osaka 565-0871}
\affiliation{Pacific Northwest National Laboratory, Richland, Washington 99352}
\affiliation{Panjab University, Chandigarh 160014}
\affiliation{Peking University, Beijing 100871}
\affiliation{University of Pittsburgh, Pittsburgh, Pennsylvania 15260}
\affiliation{Punjab Agricultural University, Ludhiana 141004}
\affiliation{Research Center for Electron Photon Science, Tohoku University, Sendai 980-8578}
\affiliation{Research Center for Nuclear Physics, Osaka University, Osaka 567-0047}
\affiliation{Theoretical Research Division, Nishina Center, RIKEN, Saitama 351-0198}
\affiliation{RIKEN BNL Research Center, Upton, New York 11973}
\affiliation{Saga University, Saga 840-8502}
\affiliation{University of Science and Technology of China, Hefei 230026}
\affiliation{Seoul National University, Seoul 151-742}
\affiliation{Shinshu University, Nagano 390-8621}
\affiliation{Showa Pharmaceutical University, Tokyo 194-8543}
\affiliation{Soongsil University, Seoul 156-743}
\affiliation{University of South Carolina, Columbia, South Carolina 29208}
\affiliation{Stefan Meyer Institute for Subatomic Physics, Vienna 1090}
\affiliation{Sungkyunkwan University, Suwon 440-746}
\affiliation{School of Physics, University of Sydney, New South Wales 2006}
\affiliation{Department of Physics, Faculty of Science, University of Tabuk, Tabuk 71451}
\affiliation{Tata Institute of Fundamental Research, Mumbai 400005}
\affiliation{Excellence Cluster Universe, Technische Universit\"at M\"unchen, 85748 Garching}
\affiliation{Department of Physics, Technische Universit\"at M\"unchen, 85748 Garching}
\affiliation{Toho University, Funabashi 274-8510}
\affiliation{Tohoku Gakuin University, Tagajo 985-8537}
\affiliation{Department of Physics, Tohoku University, Sendai 980-8578}
\affiliation{Earthquake Research Institute, University of Tokyo, Tokyo 113-0032}
\affiliation{Department of Physics, University of Tokyo, Tokyo 113-0033}
\affiliation{Tokyo Institute of Technology, Tokyo 152-8550}
\affiliation{Tokyo Metropolitan University, Tokyo 192-0397}
\affiliation{Tokyo University of Agriculture and Technology, Tokyo 184-8588}
\affiliation{University of Torino, 10124 Torino}
\affiliation{Toyama National College of Maritime Technology, Toyama 933-0293}
\affiliation{Utkal University, Bhubaneswar 751004}
\affiliation{Virginia Polytechnic Institute and State University, Blacksburg, Virginia 24061}
\affiliation{Wayne State University, Detroit, Michigan 48202}
\affiliation{Yamagata University, Yamagata 990-8560}
\affiliation{Yonsei University, Seoul 120-749}
  \author{A.~Abdesselam}\affiliation{Department of Physics, Faculty of Science, University of Tabuk, Tabuk 71451} 
  \author{I.~Adachi}\affiliation{High Energy Accelerator Research Organization (KEK), Tsukuba 305-0801}\affiliation{SOKENDAI (The Graduate University for Advanced Studies), Hayama 240-0193} 
  \author{K.~Adamczyk}\affiliation{H. Niewodniczanski Institute of Nuclear Physics, Krakow 31-342} 
  \author{H.~Aihara}\affiliation{Department of Physics, University of Tokyo, Tokyo 113-0033} 
  \author{S.~Al~Said}\affiliation{Department of Physics, Faculty of Science, University of Tabuk, Tabuk 71451}\affiliation{Department of Physics, Faculty of Science, King Abdulaziz University, Jeddah 21589} 
  \author{K.~Arinstein}\affiliation{Budker Institute of Nuclear Physics SB RAS, Novosibirsk 630090}\affiliation{Novosibirsk State University, Novosibirsk 630090} 
  \author{Y.~Arita}\affiliation{Graduate School of Science, Nagoya University, Nagoya 464-8602} 
  \author{D.~M.~Asner}\affiliation{Pacific Northwest National Laboratory, Richland, Washington 99352} 
  \author{T.~Aso}\affiliation{Toyama National College of Maritime Technology, Toyama 933-0293} 
  \author{H.~Atmacan}\affiliation{University of South Carolina, Columbia, South Carolina 29208} 
% ...
\collaboration{The Belle Collaboration}

\noaffiliation

\begin{abstract}
The precise determination of the CKM matrix element $\left| V_{cb}\right|$ is important for carrying out tests of the flavour sector of the Standard Model. In this article we present a preliminary analysis of the $\bar B^0 \to D^{*\,+} \, \ell^- \, \bar \nu_\ell$ decay mode and its isospin conjugate, selected in events that contain a  fully reconstructed $B$-meson, using 772 million $e^+ \, e^- \to \Upsilon(4S) \to B \bar B$ events recorded by the Belle detector at KEKB. Unfolded differential decay rates of four kinematic variables fully describing the $\bar B^0 \to D^{*\,+} \, \ell^- \, \bar \nu_\ell$  decay in the $B$-meson rest frame  are presented. We measure the total branching fraction $\mathcal{B}( \bar B^0 \to D^{*\,+} \, \ell^- \, \bar \nu_\ell ) = \left(4.95 \pm 0.11 \pm 0.22 \right) \times 10^{-2}$, where the errors are statistical and systematic respectively. The value of $\left|V_{cb} \right|$ is determined to be \mbox{$\left( 37.4 \pm 1.3 \right) \times 10^{-3}$}. Both results are in good agreement with current world averages.
\vspace{5ex}

Note: This version contains a corrected value for $\left| V_{cb} \right|$ and the form factor in Tables V and VI with respect to the version of Feb 6, 2017

\end{abstract}


\maketitle

\tighten

{\renewcommand{\thefootnote}{\fnsymbol{footnote}}}
\setcounter{footnote}{0}

\section{Introduction}\label{sec:intro}

Precise determinations of the values of matrix elements of the Cabibbo-Kobayashi-Maskawa (CKM) matrix~\cite{Kobayashi:1973fv,Cabibbo:1963yz} are important for testing the Standard Model of particle physics (SM).  
In this article a precise determination of the magnitude of the CKM matrix element $\left| V_{cb}\right|$ is reported, based on a measurement of the exclusive decay of $\bar B^0 \to D^{*\,+} \, \ell^- \, \bar \nu_\ell$ with $D^{*\,+} \to D^0 \pi^+$ and $D^{*\,+} \to D^+ \pi^0$ and its isospin conjugate decay mode. In addition, the unfolded differential decay rates of four kinematic quantities, described in section~\ref{sec:sltheory}, that fully characterize the semileptonic decay, are reported for the first time in this decay mode. These measurements will allow for extractions of $\left| V_{cb}\right|$ using unquenched lattice QCD calculations of the $\bar B \to D^{*}$ transition form factors beyond zero recoil when they are available in the future. This measurement complements the previous Belle untagged result in Ref.~\cite{Dungel:2010uk}, by studying the properties of the $\bar B^0 \to D^{*\,+} \, \ell^- \, \bar \nu_\ell$ decay using an orthogonal data set: the second $B$-meson in the collision is reconstructed using a fully reconstructed $B$ sample. This high purity sample allows for more precise reconstruction of the decay kinematics, at the cost of lower efficiency.  Other recent measurements of $\left| V_{cb} \right|$ using the exclusive $\bar B \to D^{*} \, \ell \, \bar \nu_\ell$ decay have been performed by the Babar experiment~\cite{Aubert:2007qs,Aubert:2007rs,Aubert:2008yv}.

This paper is organized as follows: section~\ref{sec:sltheory} briefly reviews the theory describing semileptonic \mbox{$\bar B^0 \to D^{*\,+} \, \ell^- \, \bar \nu_\ell$} decays. Section~\ref{sec:belle} provides a brief overview of the Belle detector and the data sets used in this analysis.  The event reconstruction and selection criteria are summarized in section~\ref{sec:evtreco}, while section~\ref{sec:signalext} provides an overview of the extraction of the inclusive and differential signal yields. Section~\ref{sec:unfolding} discusses the unfolding procedure. Section~\ref{sec:syst} reviews the dominant sources of systematic uncertainty. Section~\ref{sec:vcb} describes the procedure for  extracting the CKM matrix element $\left| V_{cb} \right|$. Section~\ref{sec:summary} concludes the article, with a brief summary of the key results.

\begin{figure}[h]
\includegraphics[width=0.5\linewidth,page=1,trim={0cm 0cm 0.5cm 0cm},clip]{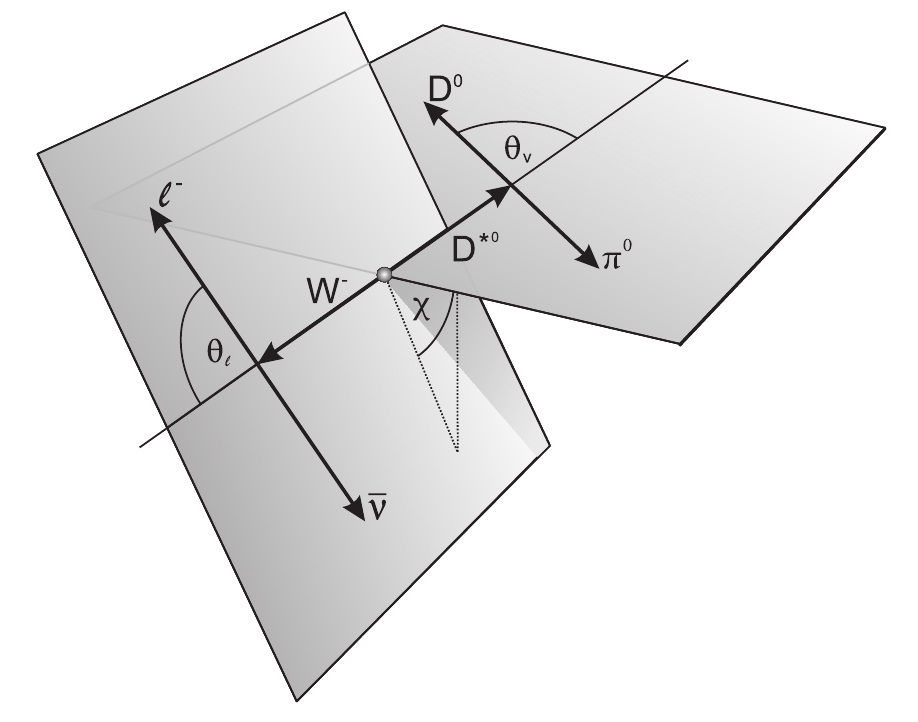}
\caption{ The helicity angles $\theta_\ell$, $\theta_v$, and $\chi$ that characterize the $\bar B \to D^{*} \, \ell \, \bar \nu_\ell$ decay are shown: the helicity angle $\theta_\ell$ is defined as the angle between the lepton and the direction opposite the $\bar B$-meson in the virtual $W$-boson rest frame; similarly $\theta_v$ is defined as the angle between the $D$ meson and the direction opposite the $\bar B$-meson in the $D^*$ rest frame; finally the angle $\chi$ is defined as the tilting angle between the two decay planes spanned by the $W-\ell$ and $D^*-D$ systems in the $\bar B$-meson rest frame. }
\label{fig:angles}
\end{figure}

\section{Theory of $\bar B \to D^{*} \, \ell^- \, \bar \nu_\ell$ decays }\label{sec:sltheory}


The $\bar B \to D^{*} \, \ell \, \bar \nu_\ell$ decay amplitude depends on one non-perturbative hadronic matrix element that can be expressed using Lorentz invariance and the equation of motion in terms of $\bar B \to D^{*}$ form factors. The four transition form factors $V$, $A_{0/1/2}$ fully describing the $\bar B \to D^{*}$ decay are defined by the hadronic current~\cite{RichmanBurchat}:
\begin{widetext}
\begin{align}\label{eq:ff}
  \langle D^{*} (p_{D^*}) | \bar c \, \gamma_\mu \, (1 - \gamma_5) \, P_L \, b |\bar  B(p_B) \rangle = &  \frac{2 i V(q^2)}{m_B + m_{D^*}} \epsilon_{\mu\nu\alpha\beta} \, \epsilon^{* \, \nu} \, p_B^\alpha \, p_{D^*}^{\beta} - (m+m_{D^*}) A_1(q^2) \left( \epsilon_\mu^* - \frac{\epsilon^* \cdot q}{q^2} q_\mu \right) \nonumber \\
  & + A_2(q^2) \, \frac{\epsilon^* \cdot q}{m_B + m_{D^*}} \left( ( p_B + p_{D^*})_\mu - \frac{m^2 - m_{D^*}^2}{q^2} q_\mu  \right) \nonumber \\
   & - 2 m_{D^*} A_0(q^2) \, \frac{\epsilon^* \cdot q}{q^2} q_\mu \, ,
  \end{align}
  where $q^\mu = \left( p_B - p_{D^*} \right)^\mu$ is the difference between the $\bar B$-meson and $D^*$-meson four momenta, and $m_B$ and $m_{D^*}$ denote the $B$-meson and $D^*$-meson masses, respectively. The $\epsilon^*$ terms denote the polarization of the $D^*$-meson. The form factors in Eq.~\ref{eq:ff} are functions of the four-momentum transfer squared $q^2$, and the differential decay rate $\bar B \to D^{*} (\to D \pi) \, \ell \, \bar \nu_\ell$ may be expressed in the zero lepton mass limit in terms of three helicity amplitudes $H_0$, $H_\pm$~\cite{RichmanBurchat}:
\begin{align} \label{eq:rate}
\frac{\text{d} \Gamma( \bar B \to D^{*} (\to D \pi) \, \ell \, \bar \nu_\ell)}{\text{d} w \, \text{d}\cos\theta_v \, \text{d}\cos\theta_{\ell} \, \text{d}\chi} =& \frac{6m_Bm_{D^*}^2}{8(4\pi)^4}\sqrt{w^2-1}(1-2 \, w\, r+r^2)\, G_F^2 \, \left|V_{cb}\right|^2 \, \times \mathcal{B}(D^{*} \to D \pi) \nonumber\\
& \times~ \biggl( (1-\cos\theta_{\ell})^2\sin^2\theta_vH_+^2 + (1+\cos\theta_{\ell})^2\sin^2\theta_vH_-^2 \nonumber \\
 &~~\qquad+ 4\sin^2\theta_{\ell}\cos^2\theta_vH_0^2 - 2\sin^2\theta_{\ell}\sin^2\theta_v\cos2\chi H_+H_-\nonumber \\
 &~~\qquad- 4\sin\theta_{\ell}(1-\cos\theta_{\ell})\sin\theta_v\cos\theta_v\cos\chi H_+H_0\nonumber \\
 &~~\qquad+ 4\sin\theta_{\ell}(1+\cos\theta_{\ell})\sin\theta_v\cos\theta_v\cos\chi H_-H_0 \biggr) ~,
 \end{align}
where the $q^2$ is written as  the product of the four-velocities of the initial- and final-state meson, \mbox{$w = \left( m_B^2 + m_{D^*}^2 - q^2 \right)/(2 m_B m_{D^*})$} for later convenience and $r = m_{D^*} / m_B$. 
 The helicity amplitudes are related to the form factors as
\begin{align}
  H_{\pm}  &= \left(m_B + m_{D^*} \right) A_1(q^2) \mp \frac{2 m_B}{m_B + m_{D^*}} \left| p_{D^*} \right| V(q^2) \, , \\
  H_0  & = \frac{1}{2 m_{D^*} \sqrt{q^2}} \bigg( \left(m_B^2 - m_{D^*}^2 - q^2 \right) \left( m_B + m_{D^*} \right) A_1(q^2) 
 - \frac{4 m_B^2 \left| p_{D^*} \right|^2}{m_B + m_{D^*}} A_2(q^2)  \bigg) \, .
\end{align}
\end{widetext}
The light constituents of the $\bar B$- and $D^*$-mesons are only lightly perturbed if the velocities of the $b$- and $c$-quarks inside the $\bar B$- and $D^*$-mesons are similar, e.g. for $q^2 = q^2_{\rm max}$ or $w \sim 1$ ~\cite{WiseManohar}. 
 The four form factors in Eq.~\ref{eq:ff} can be expressed in terms of a single universal form factor $h_{A_1}(w)$ and three ratios $R_i(w)$,
\begin{align}
 A_1  = \frac{w+1}{2} r' h_{A_1}(w) \, , \qquad  A_0  =  \frac{R_0(w)}{r'} h_{A_1}(w) \, , \nonumber \\
 A_2  = \frac{R_2(w)}{r'} h_{A_1}(w)\, , \qquad  V  =  \frac{R_1(w)}{r'} h_{A_1}(w) \, ,
\end{align}
with $r' = 2 \sqrt{m_B m_{D^*}} / \left( m_B + m_{D^*} \right)$. Analyticity and unitarity impose strong constraints on heavy meson decay form factors~\cite{Grinstein:1992hq} and the universal form factor and ratios can be expressed in terms of five parameters \mbox{\{$h_{A_1}(1)$, $\rho_{D^{*}}^2$, $R_{0/1/2}(1)$\}}, cf. Ref.~\cite{Caprini:1997mu}:
\begin{align}\label{eq:ffhqet}
 h_{A_1}(w) &= h_{A_1}(1) \bigg( 1 - 8 \rho_{D^{*}}^2 z + \left(53 \rho_{D^{*}}^2 - 15 \right) z^2 - \left(231 \rho_{D^{*}}^2 - 91  \right) z^3  \bigg) \, ,  \\
 R_0(w) &= R_0(1) - 0.11(w - 1) + 0.01(w-1)^2\, ,  \\
 R_1(w) &= R_1(1) - 0.12(w - 1) + 0.05(w-1)^2\, ,  \\
 R_2(w) &= R_2(1) + 0.11(w - 1) - 0.06(w-1)^2\,  , 
\end{align}
with $z = \left( \sqrt{w+1} - \sqrt{2}  \right) / \left(  \sqrt{w+1} + \sqrt{2} \right)$. The ratio $R_0(w)$ is not important for decays involving light leptons. 
The current state-of-the-art unquenched calculation Ref.~\cite{Bailey:2014tva} uses up to three light-quark flavours and yields \mbox{$h_{A_1}(1) = 0.906 \pm 0.013$}. Equation~\ref{eq:rate} receives additional electroweak corrections that can be introduced by the replacement of \mbox{$h_{A_1}(1) \to h_{A_1}(1) \, \eta_{EW}$} with $\eta_{EW} = 1.0066$ from Ref.~\cite{Sirlin}. The remaining three parameters, \mbox{$\{\rho_{D^{*}}^2$, $R_{1/2}(1)\}$}, need to be determined experimentally by analyzing the differential $\bar B \to D^{*} \, \ell \, \bar \nu_\ell$ spectrum to convert the measured branching fraction into a value of $\left| V_{cb}  \right|$: 
\begin{align}
 \left| V_{cb} \right| & = \sqrt{ \frac{ \mathcal{B}(\bar B \to D^{*} \, \ell \, \bar \nu_\ell) }{ \tau \, \Gamma(\bar B \to D^{*} \, \ell \, \bar \nu_\ell)} } \, ,
\end{align}
where  $\tau$ is the $B$-meson lifetime and  $\Gamma(\bar B \to D^{*} \, \ell \, \bar \nu_\ell)$ is the decay rate with the CKM factor omitted.

\section{The Belle detector and data set}\label{sec:belle}

The data sample used in this measurement was recorded with the Belle detector~\cite{unknown:2000cg}, that operated at the KEKB storage ring~\cite{Kurokawa:2001nw} between 1999 and 2010. This analysis uses an integrated luminosity of \mbox{711 fb${}^{-1}$} recorded at the centre-of-mass energy of \mbox{$\sqrt{s} = 10.58$ GeV}, corresponding to 772 million $e^+ e^- \to \Upsilon(4S) \to B \bar B$ events. KEKB is an asymmetric $e^+e^-$ collider in which the centre-of-mass of the colliding beams moves with a velocity of $\beta = 0.425$ along the beam axis in the laboratory rest frame. 
 The Belle detector is a large solid angle magnetic spectrometer optimized to reconstruct $e^+ e^- \to \Upsilon(4S) \to B \bar B$ collisions. Its principal detector components are: the silicon vertex detector, the 50-layer central drift chamber, the array of aerogel based Cherenkov counters, the time-of-flight scintillation counters, and the electromagnetic calorimeter built from CsI(Tl) crystals located inside a superconducting solenoid coil producing a 1.5 T magnetic field. The outer layer consists of an instrumented iron flux-return allowing the identification of $K_L^0$ mesons and muons. During data taking two different inner detector configurations were used: the first configuration,  corresponding to  152 million $B \bar B$ pairs, consisted of a 2.0 cm beampipe and a 3-layer silicon vertex detector. The second configuration,  used to record the remaining 620 million $B \bar B$ pairs, consisted of a 1.5 cm beampipe, a 4-layer silicon vertex detector, and a small-cell inner drift chamber~\cite{svd2}. A more detailed description of the detector and its performance can be found in Refs.~\cite{unknown:2000cg,svd2}.

Simulated Monte Carlo (MC) events are used to evaluate background contamination, reconstruction efficiency and acceptance, and for the unfolding procedure. The samples were generated using the \texttt{EvtGen} generator~\cite{Lange:2001uf}, with event sizes corresponding to approximatively ten times that of the Belle collision data. The interaction of particles traversing the detectors is simulated using  \texttt{GEANT3}~\cite{brun1984geant}.  QED final state radiation was  simulated using  \texttt{PHOTOS}~\cite{barberio1994photos}. The form factor parametrization in section~\ref{sec:sltheory} is used to model the semileptonic $\bar B \to D^{*} \, \ell \, \bar \nu_\ell$ signal. The  $\bar B \to D \, \ell \, \bar \nu_\ell$ decays are modelled using the form factor parametrization in Ref.~\cite{Caprini:1997mu}. Semileptonic decays into orbitally excited charmed mesons, $\bar B \to D^{**} \, \ell \, \bar \nu_\ell$, were modelled using the form factor parametrization of Ref.~\cite{ref:LLSW}. 
The branching fractions for $B$-meson and charm decays are taken from Ref.~\cite{pdg}. Efficiencies in the MC are corrected using data driven control samples.

\section{Event reconstruction and selection}\label{sec:evtreco}

Collision events are reconstructed using the hadronic full reconstruction algorithm of Ref.~\cite{Feindt:2011mr}: In the algorithm one of the $B$-mesons, called the $B_{\rm tag}$-candidate, is reconstructed in hadronic decay channels using over 1100 decay modes. The efficiency of this approach is approximately 0.3\% and 0.2\% for charged and neutral $B$-mesons, respectively. Despite the relatively low efficiency, knowledge of the charge and momenta of the decay constituents in combination with the known beam-energy allows one to precisely infer the flavour and four-momentum of the second $B$-meson produced in the collision. The $B_{\rm tag}$-candidates are required to have a beam constrained $B$-meson mass, $$M_{\rm bc} = \sqrt{ s/4 - \left| \vec p_{\rm tag} \right|^2} \, $$ larger than $5.265$ GeV~\footnote{We use natural units with $\bar h = c = 1$.}, where $\sqrt{s}$ denotes the centre-of-mass energy of the colliding $e^+e^-$ pair and $\vec p_{\rm tag}$ denotes the reconstructed three-momentum of the $B_{\rm tag}$-candidate in the centre-of-mass frame of the colliding $e^+e^-$ pair. In addition a requirement of $ -0.15 \, \text{GeV} < \Delta E < 0.1$ GeV is imposed with $$\Delta E = E_{\rm tag} - \sqrt{s}/2 $$ and $E_{\rm tag}$ denoting the reconstructed energy of the $B_{\rm tag}$-candidate in the centre-of-mass frame of the colliding $e^+e^-$ pair. In each event a single  $B_{\rm tag}$-candidate is chosen according to the highest classifier score of the hierarchical full reconstruction algorithm. All tracks and neutral clusters used to form the $B_{\rm tag}$-candidate are removed from the event to define a signal side.

\subsection{Signal side Reconstruction}
The signal $\bar B^0 \to D^{*\,+} \, \ell^- \, \bar \nu_\ell$ decay is reconstructed in three steps~\footnote{Isospin conjugated modes are implied throughout the manuscript.}:
\begin{enumerate}
\item A lepton candidate (an electron or muon) is reconstructed, and identified using a particle identification (PID) likelihood ratio described in Ref.~\cite{unknown:2000cg}.  
A minimal lepton momentum of $0.3$ GeV for electrons and $0.6$ GeV for muons is required, while the track of the lepton candidate must be within the detector acceptance with a polar angle relative to the beam axis of $17^{\circ} < \theta_e < 150^{\circ}$ and  $ 25^{\circ} < \theta_\mu < 145^{\circ}$  for electrons and muons, respectively. In addition, impact parameter requirements on the lepton candidates in the plane perpendicular to the beam are applied. For electron candidates, bremsstrahlung and final state radiation photons are recovered using a cone around the lepton trajectory with an opening angle of $5^{\circ}$. In the case that several photon candidates are in this cone, the one with the smallest opening angle to the electron is used. Events with more than one well identified lepton are vetoed. 

\item Charged and neutral $D$-meson candidates are reconstructed from kaon candidates, charged tracks and $\pi^0$ candidates. Kaons and pions are identified as described in Ref.~\cite{unknown:2000cg} using a PID likelihood ratio, and must also satisfy impact parameter requirements. The $\pi^0$ candidates are reconstructed from photon candidates, which consist of clusters in the calorimeter not matched to any track. The energy requirement for photon candidates evolves as a function of polar angle: $E_\gamma > 100$ MeV for $\theta_\gamma < 33^{\circ}$, $E_\gamma > 50$ MeV for $ 33^{\circ} < \theta_\gamma < 128^{\circ}$, and $E_\gamma > $ 150 MeV for $\theta_\gamma > 128^{\circ}$. The invariant mass of the $\pi^0$ candidates must fall within a mass window of $M_{\pi^0} = [0.12,0.15)$ GeV. 
All combinations of particles that form $D^0$ or $D^+$ meson candidates with an invariant mass within 14 MeV of $m_{D^+} = 1870$ MeV and $m_{D^0} = 1865$ MeV respectively, are used in a fit for a secondary vertex to select a single $D^0$ or $D^+$ candidate per event. The decay modes used are $D^+ \to K^- \pi^+ \pi^+$,  $D^0 \to K^- \pi^+$,  $ D^0 \to K^- \pi^+ \pi^0$,  $D^0 \to K^- \pi^- \pi^+ \pi^+$, which account for 9.4\% and 26.3\% of the total $D^+$ and $D^0$ branching fractions. In events with a $D^+$ candidate no additional track is allowed on the signal side. In events with a $D^0$ candidate exactly one additional  track is required. 

\item Finally candidate $D^{*}$-mesons are reconstructed: here the decay of $D^{*\,+} \to D^0 \pi^+$ is reconstructed by combining the four-momentum of the reconstructed $D^0$ with the remaining charged track in the event. Events with $D^{*\,+} \to D^0 \pi^+$ candidates are rejected if the reconstructed mass difference $\Delta M = M_{D^*} - M_{D}$ has a value outside a window of $[135,155)$ MeV, corresponding to three times the expected $\Delta M$ resolution as estimated from MC. The decay of $D^{*\,+} \to D^+ \pi^0$ is reconstructed by combining the four-momentum of the reconstructed $D^{+}$ with all possible $\pi^0$ candidates and a single candidate is chosen by selecting the candidate with a $\Delta M = M_{D^*} - M_{D}$ closest to the expected value of $140$ MeV and fall in the window $[130,150)$ MeV, corresponding to three times the expected resolution of $\Delta M$. The $D^{*\,+} \to D^0 \pi^+$ and $D^{*\,+} \to D^+ \pi^0$ decays account for 98.4\% of the total $D^{*\,+}$ branching fraction.
\end{enumerate}

\subsection{Calibration of the hierarchical full reconstruction algorithm}\label{subsec:calibration}

The efficiency of the full hadronic reconstruction algorithm is calibrated using a procedure described in Ref.~\cite{Glattauer:2015teq} based on a study of inclusive $\bar B \to X \, \ell \, \bar \nu_\ell$ decays. In this approach full reconstruction events are selected by requiring exactly one lepton on the signal side, employing the same lepton and $B_{\rm tag}$ selection criteria as outlined above. The $\bar B \to X \, \ell \, \bar \nu_\ell$ enriched events are split into subsamples according to their hadronic $B_{\rm tag}$ final state topology and further separated into specific ranges of the multivariate classifier used in the hierarchical selection. Each subsample is studied individually to derive a calibration factor for the hadronic tagging efficiency: this is done by confronting the number of inclusive semileptonic $B$-meson decays, $N(\bar B \to X \, \ell \, \bar \nu_\ell)$, in data with the expectation from the simulation, $N^{MC}(\bar B \to X \, \ell \, \bar \nu_\ell)$, assuming the branching fraction of Ref.~\cite{pdg}. The semileptonic yield is determined by a binned likelihood fit to the spectrum of the lepton three-momentum and the correction factors in each subsample is given by 
\begin{align}
 C_{\rm tag} = N(\bar B \to X \, \ell \, \bar \nu_\ell) / N^{MC}(\bar B \to X \, \ell \, \bar \nu_\ell) \, .
\end{align} 
The free parameters of the fit were prompt semileptonic $\bar B \to X \, \ell \, \bar \nu_\ell$ decays, fake lepton contributions and secondary true lepton contributions and in total 1120 correction factors were determined. The largest uncertainties on the $C_{\rm tag}$ correction factors are from the assumed $\bar B \to X \, \ell \, \bar \nu_\ell$ shape and the lepton PID performance, cf. Section~\ref{sec:syst}.

\section{Reconstruction of kinematic quantities and signal extraction} \label{sec:signalext}

\begin{figure}[t]
\vspace{0.5ex}
\includegraphics[width=0.8\linewidth,page=1,trim={0cm 0cm 0.5cm 0cm},clip]{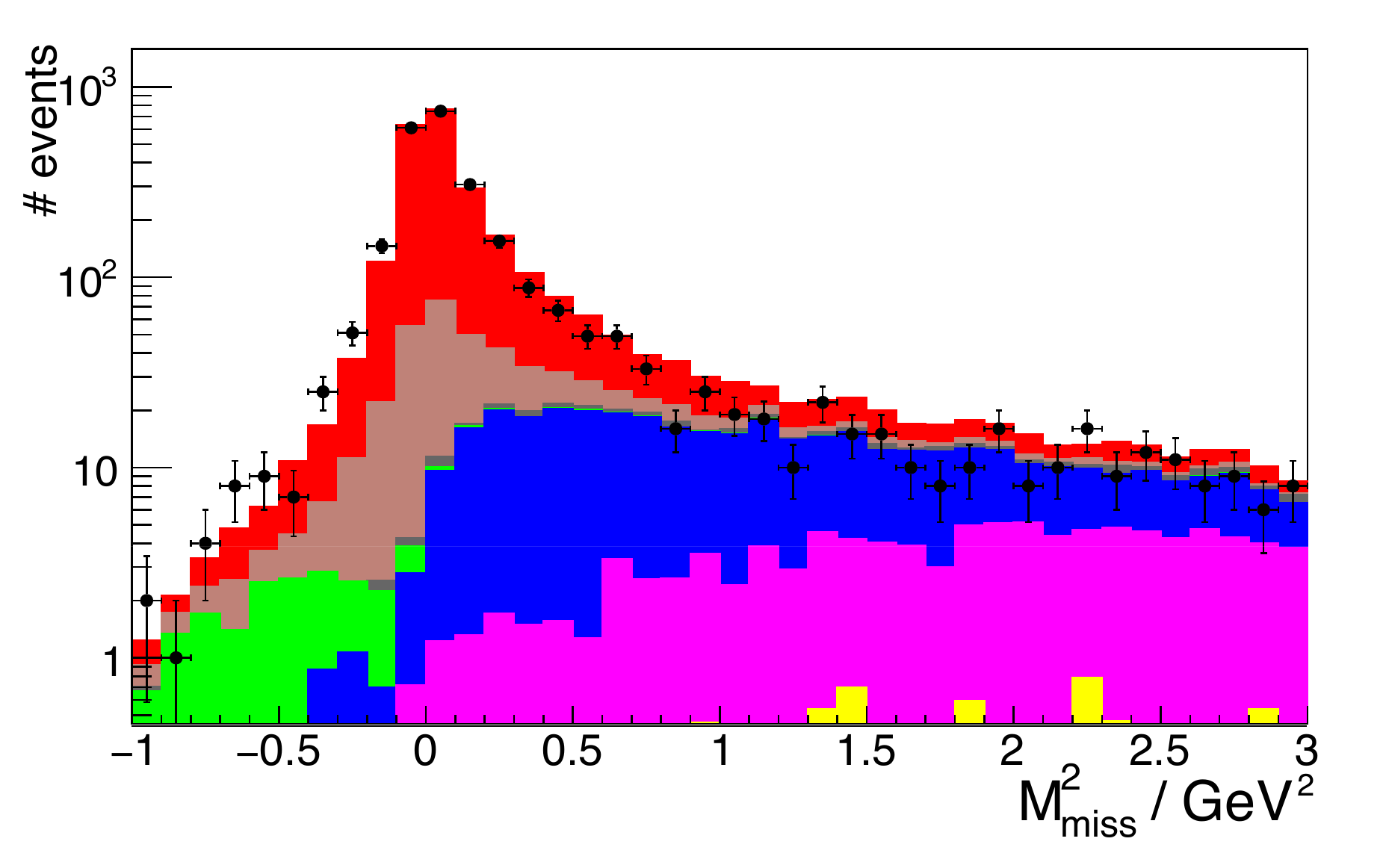} 
\put(-130.,130.){\includegraphics[width=0.26\linewidth,page=1,clip]{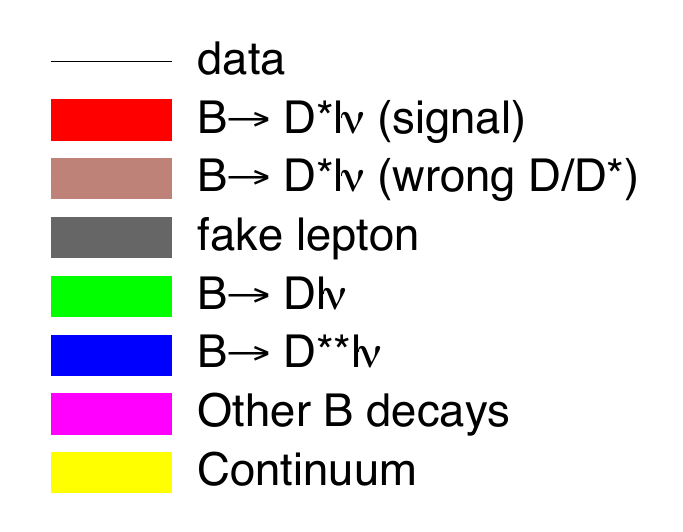}}
\vspace{-3ex}
\caption{ The $M_{\rm miss}^2$ distribution of all events after the $\bar B^0 \to D^{*\,+} \, \ell^- \, \bar \nu_\ell$ reconstruction. The coloured histograms correspond to either correctly (red) or incorrectely reconstructed signal (brown) or various backgrounds. The largest background comes from semileptonic $\bar B \to D^{**} \, \ell \, \bar \nu_\ell$ decays and other $B$-meson decays.
}
\label{fig:B0Incl}
\end{figure}

The signal $\bar B^0 \to D^{*\,+} \, \ell^- \, \bar \nu_\ell$  can be reconstructed using the missing momentum in the collision,
\begin{align}
p_{\rm miss} & = p_\nu  = p_{e^+ e^-} - p_{\rm tag} - p_{D^*} - p_\ell \, ,
\end{align}
where the subscript indicates the corresponding four-momenta of the colliding $e^+e^-$ pair, the tag side $B$-meson, and the reconstructed signal side $D^{*}$ and lepton. To separate signal $\bar B^0 \to D^{*\,+} \, \ell^- \, \bar \nu_\ell$ decays from background processes, the missing mass squared used, calculated from the missing momentum via
\begin{align}
 M_{\rm miss}^2 = p_{\rm miss}^2 \, .
\end{align}
Only correctly reconstructed signal peaks at $M_{\rm miss}^2=0$, consistent with a single missing neutrino. 
Figure~\ref{fig:B0Incl} shows the reconstructed $M_{\rm miss}^2$ distribution after the initial selection and the reconstruction of the $D^{*\, +}$-meson: correctly reconstructed $\bar B^0 \to D^{*\,+} \, \ell^- \, \bar \nu_\ell$ signal decays are shown in red and sharply peak around $M_{\rm miss}^2 \sim 0$. Decays of real $D^0$, $D^+$ or $D^{*\, +}$ candidates that have been incorrectly reconstructed are shown in brown and exhibit very similar resolution in $M_{\rm miss}^2$. Fake lepton contributions, continuum events and $\bar B \to D \, \ell \, \bar \nu_\ell$ decays are negligible; the largest selected background contribution is from $\bar B \to D^{**} \, \ell \, \bar \nu_\ell$ decays and other non-semileptonic $B$-meson decays that pass the selection criteria. Most of these are from cascade decays, where a secondary decay of a $D$-meson produced a lepton. 
\begin{figure}[h!]
\includegraphics[width=0.41\textwidth,page=1,trim={0cm 0cm 0.5cm 0cm},clip]{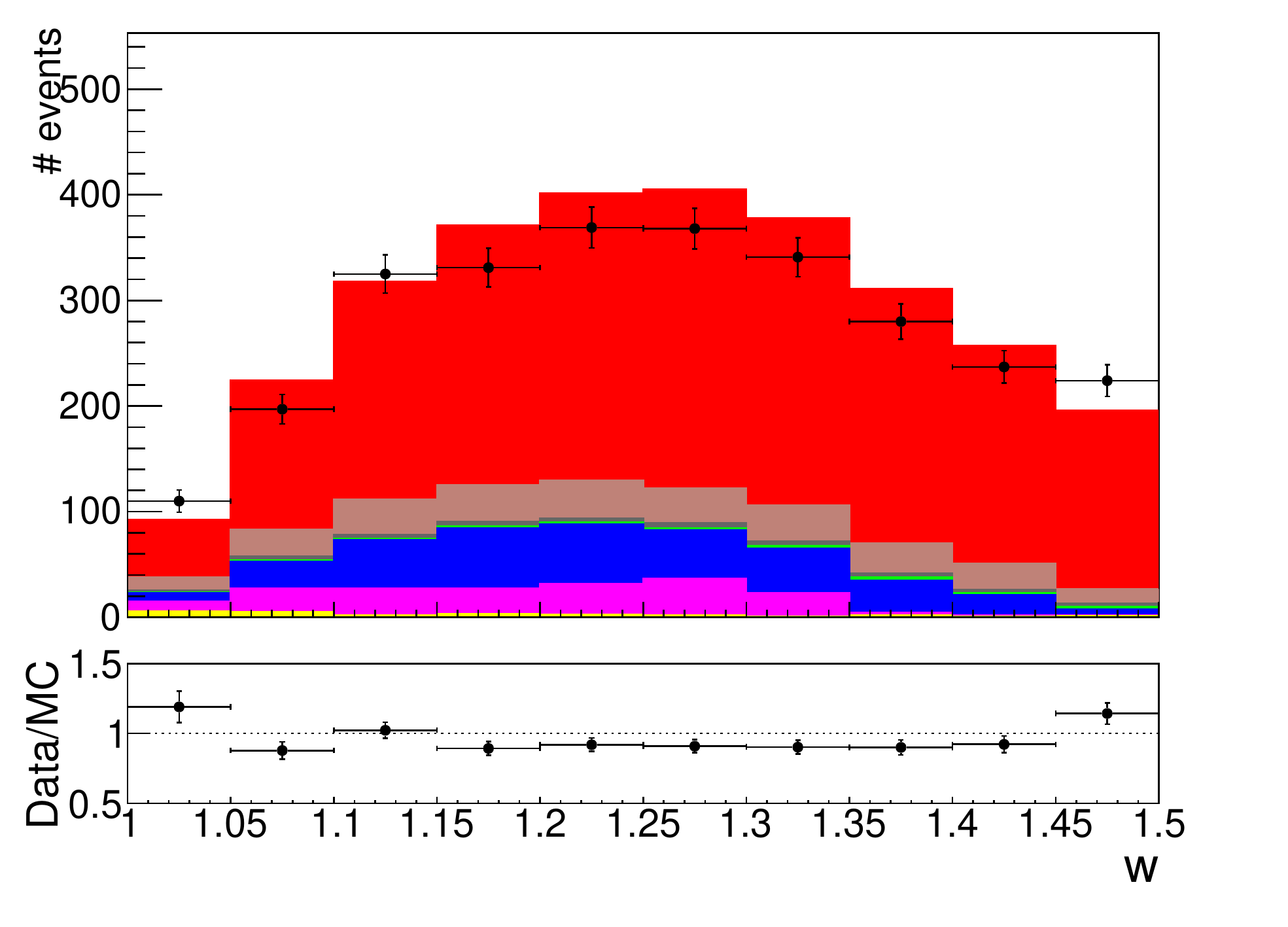} 
\includegraphics[width=0.44\textwidth,page=1,trim={0cm 0cm 0.5cm 0cm},clip]{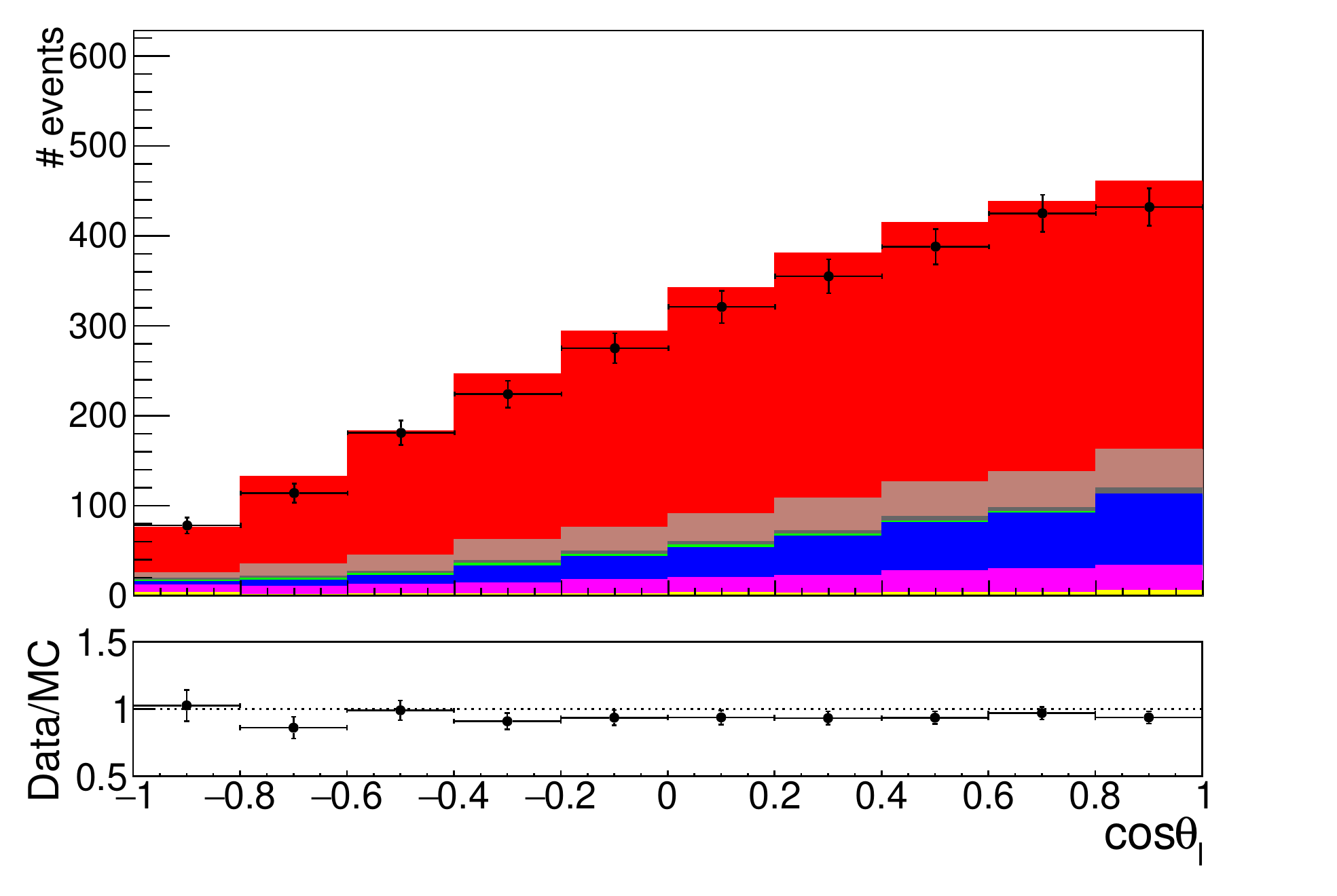} 
\put(-177.,88.){\includegraphics[width=0.13\linewidth,page=1,clip]{./figures/legend.pdf}} \\
\includegraphics[width=0.43\textwidth,page=1,trim={0cm 0cm 0.5cm 0cm},clip]{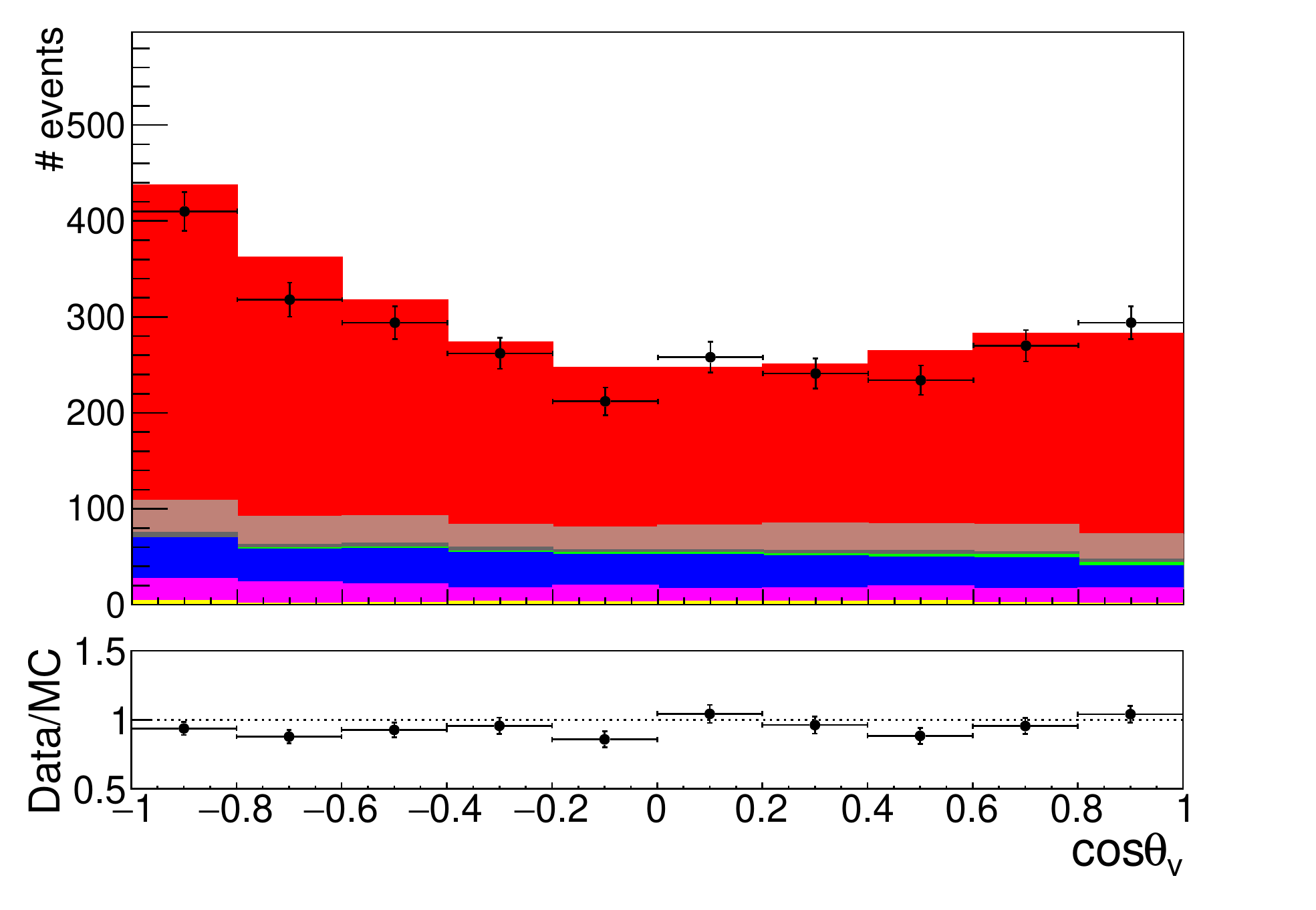} 
\includegraphics[width=0.41\textwidth,page=1,trim={0cm 0cm 0.5cm 0cm},clip]{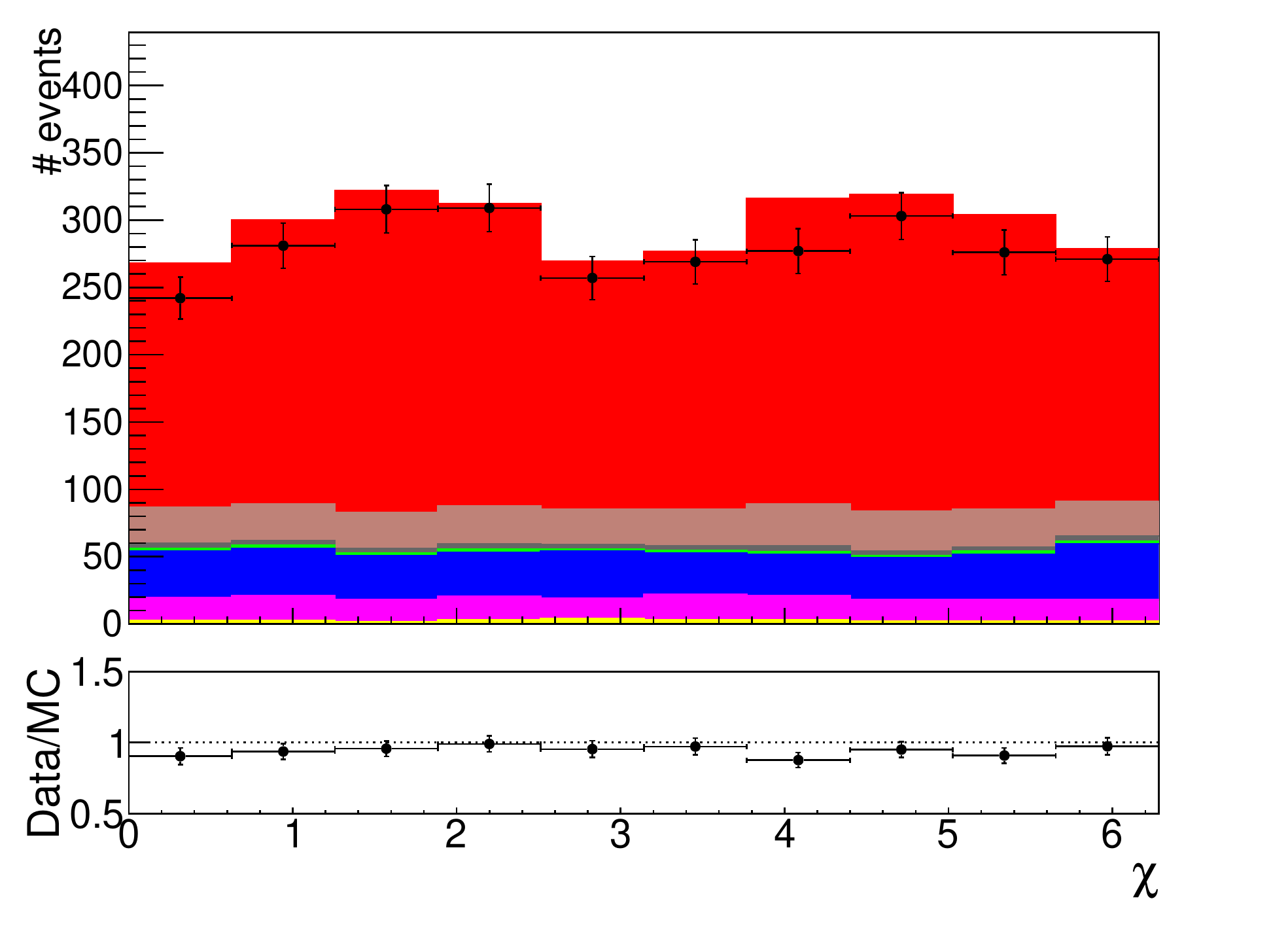} 
\caption{ 
The reconstructed kinematic variables $w$, $\cos \theta_\ell$, $\cos \theta_v$, and $\chi$ are shown, as defined in the text.
}
\label{fig:B0Vars}
\end{figure}

The kinematic variables $w$, $\cos \theta_\ell$, $\cos \theta_v$ and $\chi$ are reconstructed from the four momenta of the signal side $D^{*\, +}$, the charged lepton, and the tag-side $B$-meson. The hadronic recoil, $w$, is determined by reconstructing the four-momentum of the signal-side $\bar B$-meson as $p_B = p_{e^+ e^-} - p_{\rm tag}$ and combining it with the $D^{*\,+}$ four-momentum; the decay angles are calculated from all four-vectors boosted into the rest-frame of the signal $\bar B$-meson. The helicity angle $\theta_\ell$ is the angle between the lepton and the direction opposite to the $\bar B$-meson in the virtual $W$-boson rest frame. The helicity angle $\theta_v$ is  the angle between the $D$ meson and the direction opposite the $\bar B$-meson in the $D^*$ rest frame. Finally, $\chi$ is the angle between the two decay planes spanned by the $W-\ell$ and $D^*-D$ systems in the $\bar B$-meson rest frame. Figure~\ref{fig:B0Vars} compares the reconstructed kinematic variables in data with the expectation from MC. 

The number of $\bar B^0 \to D^{*\,+} \, \ell^- \, \bar \nu_\ell$ signal events is calculated using an unbinned maximum likelihood fit to the $M_{\rm miss}^2$ distribution. 
Incorrectly reconstructed $D^{*\,+}$-mesons are treated as a resolution effect in the variables in question when extracting the form factors in Section~\ref{sec:vcb}. Similarly, all backgrounds are merged into a single component, fixing their relative contributions to the values in the simulation. The likelihood function has the form 
\begin{align}\label{eq:likelihood}
 \mathcal{L}(M_{\rm miss}^2; \nu^{\rm sig}, \nu^{\rm bkg}) & =  \frac{e^{- \nu}}{n!} \prod_i^{n} \bigg( \nu^{\rm sig} \mathcal{S}(M_{{\rm miss} \, i}^2) + \nu^{\rm bkg} \mathcal{B}(M_{{\rm miss} \, i}^2)   \bigg) \,
\end{align}
where $\nu^{\rm sig}$ is the fitted number of signal events,  $\nu^{\rm bkg}$ is the fitted number of background events, and $\nu = \nu^{\rm sig} + \nu^{\rm bkg}$ is the mean value of the Poisson distribution  for $n$ observed events in data. The terms $\mathcal{S}(M_{{\rm miss} \, i}^2)$ and $\mathcal{B}(M_{{\rm miss} \, i}^2)$ denote the signal and background probability distribution functions (PDFs) respectively, evaluated for an event $i$ with a value of missing mass squared of $M_{{\rm miss} \, i}^2$. The likelihood Eq.~\ref{eq:likelihood} is maximized numerically, either for all events or in bins of the kinematic observables. The number of signal events is not constrained to be positive in the fit. The signal and background PDFs are constructed from signal and background MC events using Gaussian kernel estimators~\cite{Cranmer:2000du} and the fit tested with pseudo-experiments and independent subsets of MC events to ensure the the procedure is statistically unbiased. 

\subsection{Total branching fraction fit result}

The number of signal events obtained from the fit is $\nu^{\rm sig} = 2374 \pm 53$. We also provide separate results for electron and muon final states, which are in good agreement with the expectation from MC as summarised in Table~\ref{tab:fit_summary}. The number of signal decays can be converted into the $\bar B^0 \to D^{*\,+} \, \ell^- \, \bar \nu_\ell$ branching fraction using the total number of $B \bar B$ events produced at Belle of $N_{B \bar B} = \left(772 \pm 11\right) \times 10^6 $, the product of the reconstruction and tagging efficiency $\left( \epsilon_{\rm reco} \epsilon_{\rm tag} \right)$, and the $B^0/B^+$ production ratio $f_{+0}$ defined as
\begin{align}
f_{+0} & =  \frac{ \mathcal{B}( \Upsilon(4S) \to B^+ \bar B^+) }{ \mathcal{B}( \Upsilon(4S) \to B^0 \bar B^0)}  = 1.058 \pm 0.024 \, ,
\end{align}
from Ref.~\cite{pdg}. The product of the reconstruction and tagging efficiency is determined from MC after application of the calibration procedure described in Section~\ref{subsec:calibration}:
\begin{align}
\left( \epsilon_{\rm reco} \epsilon_{\rm tag} \right) = 3.19 \times 10^{-5} \, .
\end{align}
The measured $\bar B^0 \to D^{*\,+} \, \ell^- \, \bar \nu_\ell$ branching fraction is then given by
\begin{align}
 \mathcal{B}(\bar B^0 \to D^{*\,+} \, \ell^- \, \bar \nu_\ell) & = \frac{ \nu^{\rm sig} \left( \epsilon_{\rm reco} \epsilon_{\rm tag} \right)^{-1} }{ 4 N_{B \bar B} \left( 1 + f_{+0} \right)^{-1} } \, ,
\end{align}
where the factor of $4$ accounts for having two $B$-mesons in each decay and that we average the branching fraction over both light leptons. We measure
\begin{align}\label{res:b0bf}
 \mathcal{B}(\bar B^0 \to D^{*\,+} \, \ell^- \, \bar \nu_\ell) & = \left(4.95 \pm 0.11 \pm 0.22 \right)\times 10^{-2} \, ,
\end{align}
where the first error in the branching fraction is statistical and the second error from systematic uncertainties. A full breakdown of the systematic uncertainties is discussed in Section~\ref{sec:syst}. This branching fraction can be compared with the current world average
\begin{align}
 \mathcal{B}_{\rm wa}(\bar B^0 \to D^{*\,+} \, \ell^- \, \bar \nu_\ell) & = \left(4.88 \pm 0.01 \pm 0.10 \right)\times 10^{-2} \, ,
\end{align}
from Ref.~\cite{hfag} and we find good agreement. For the separate branching fractions to $\ell = e$ and $\ell = \mu$ we find
\begin{align}
 \mathcal{B}(\bar B^0 \to D^{*\,+ } \, e^- \,  \bar \nu_e) & = \left(5.04 \pm 0.15 \pm 0.23 \right)\times 10^{-2} \, , 
 \end{align}
 and
 \begin{align}
 \mathcal{B}(\bar B^0 \to D^{*\,+} \, \mu^- \, \bar \nu_\mu) & = \left(4.84 \pm 0.15 \pm 0.22 \right)\times 10^{-2} \, ,
\end{align}
where both are in good agreement with each other and hence with the average Eq.~\ref{res:b0bf}. The ratio of both branching fractions is measured to be
 \begin{align}
 R_{e\mu} = \frac{ \mathcal{B}(\bar B^0 \to D^{*\,+} \, e^- \, \bar \nu_e) }{  \mathcal{B}(\bar B^0 \to D^{*\,+ } \, \mu^- \, \bar \nu_\mu)}  & =  1.04 \pm 0.05 \pm 0.01 \, .
 \end{align} 

\begin{table}[t]
\begin{tabular}{c|cccccc}
\hline\hline
$\ell$ & $\nu^{\rm sig}$  &  $\nu^{\rm sig}_{\rm MC}$  & $\epsilon_{\rm reco} \epsilon_{\rm tag}$  \\ \hline 
$e+\mu$ &  $2374 \pm 53$ & $2310.1$ & $3.19 \times 10^{-5}$ \\
$e$ & $1306 \pm 40$ & $1248.8$ & $3.45 \times 10^{-5}$ \\
$\mu$ & $1066 \pm 34$ & $1061.3$ & $2.93 \times 10^{-5}$ \\
\hline\hline
\end{tabular}
\caption{The measured ($\nu^{\rm sig}$) and expected ($\nu^{\rm sig}_{\rm MC}$) \mbox{$\bar B^0 \to D^{*\,+} \, \ell^- \, \bar \nu_\ell$} signal yields are listed for the combined fit and for the electron and muon subsamples, as well as the product of the reconstruction and tagging efficiencies.}
\label{tab:fit_summary}
\end{table}

\subsection{Differential fit and statistical correlations}

\begin{figure}[h!]
\subfigure[ \, $w \in [1, 1.05)$ ]{\includegraphics[width=0.4\linewidth]{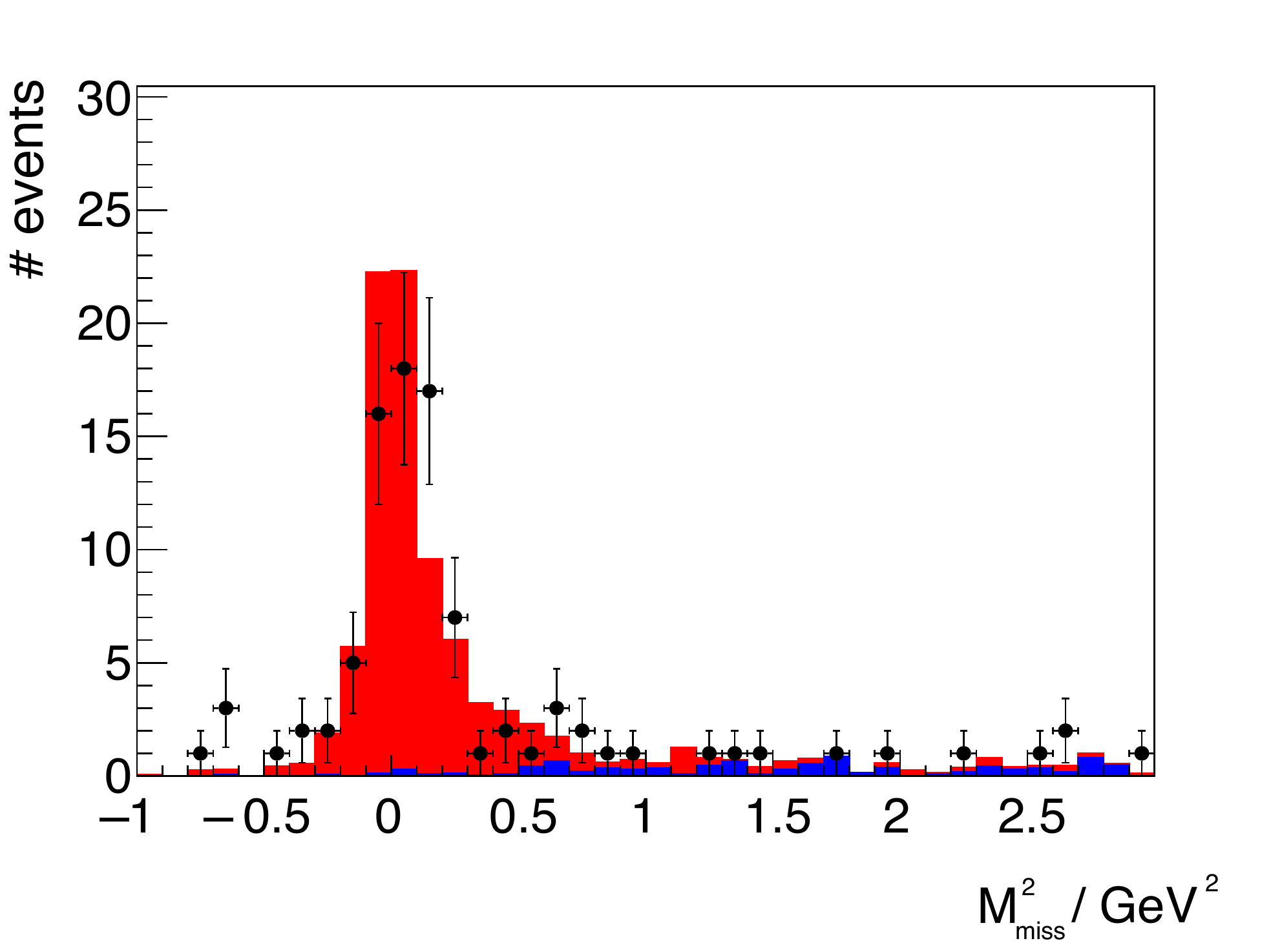}
\put(-90.,80.){\includegraphics[width=0.14\linewidth,page=1,clip]{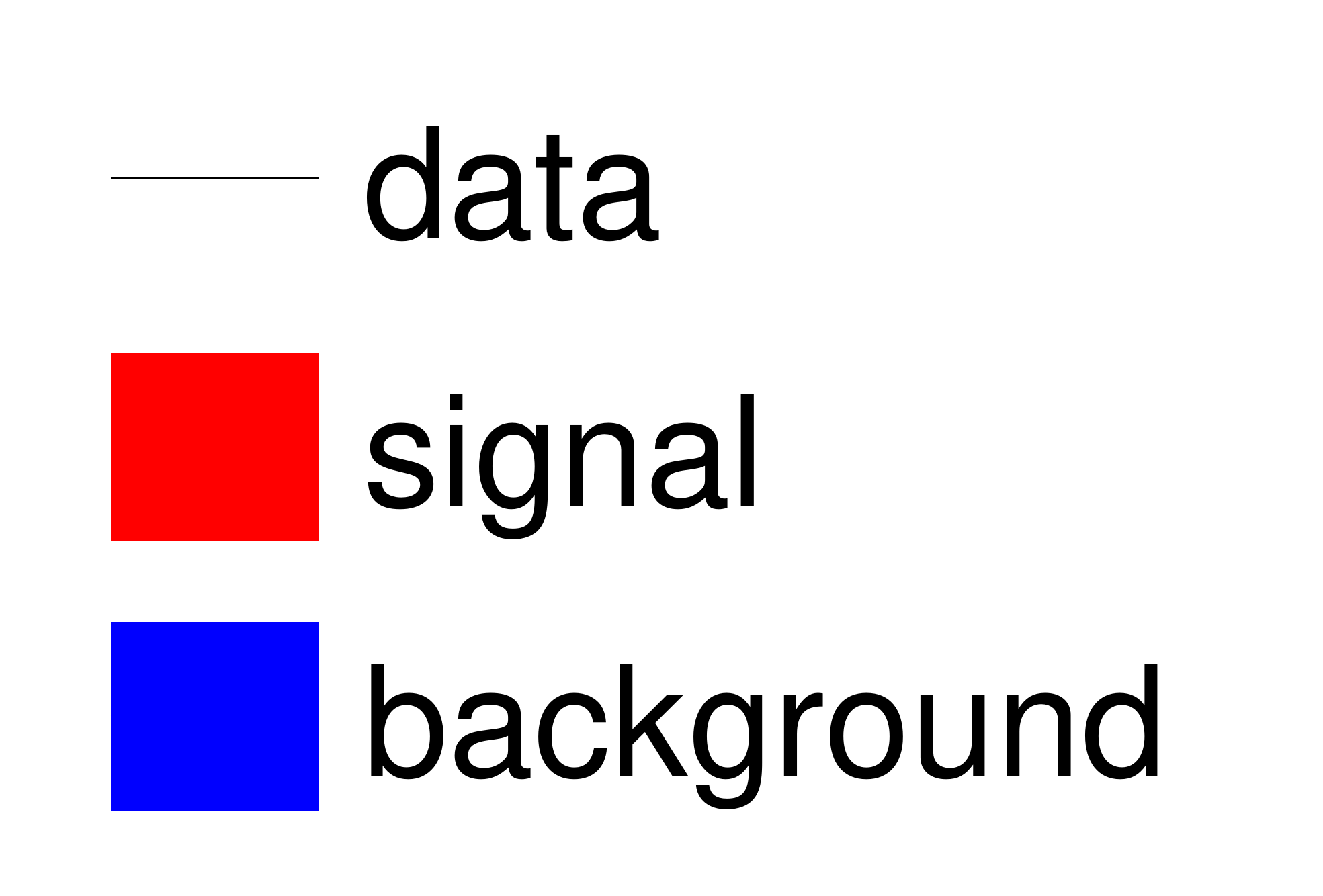}}}
\subfigure[ \, $\cos \theta_\ell \in [0.8, 1.0)$ ]{\includegraphics[width=0.4\linewidth]{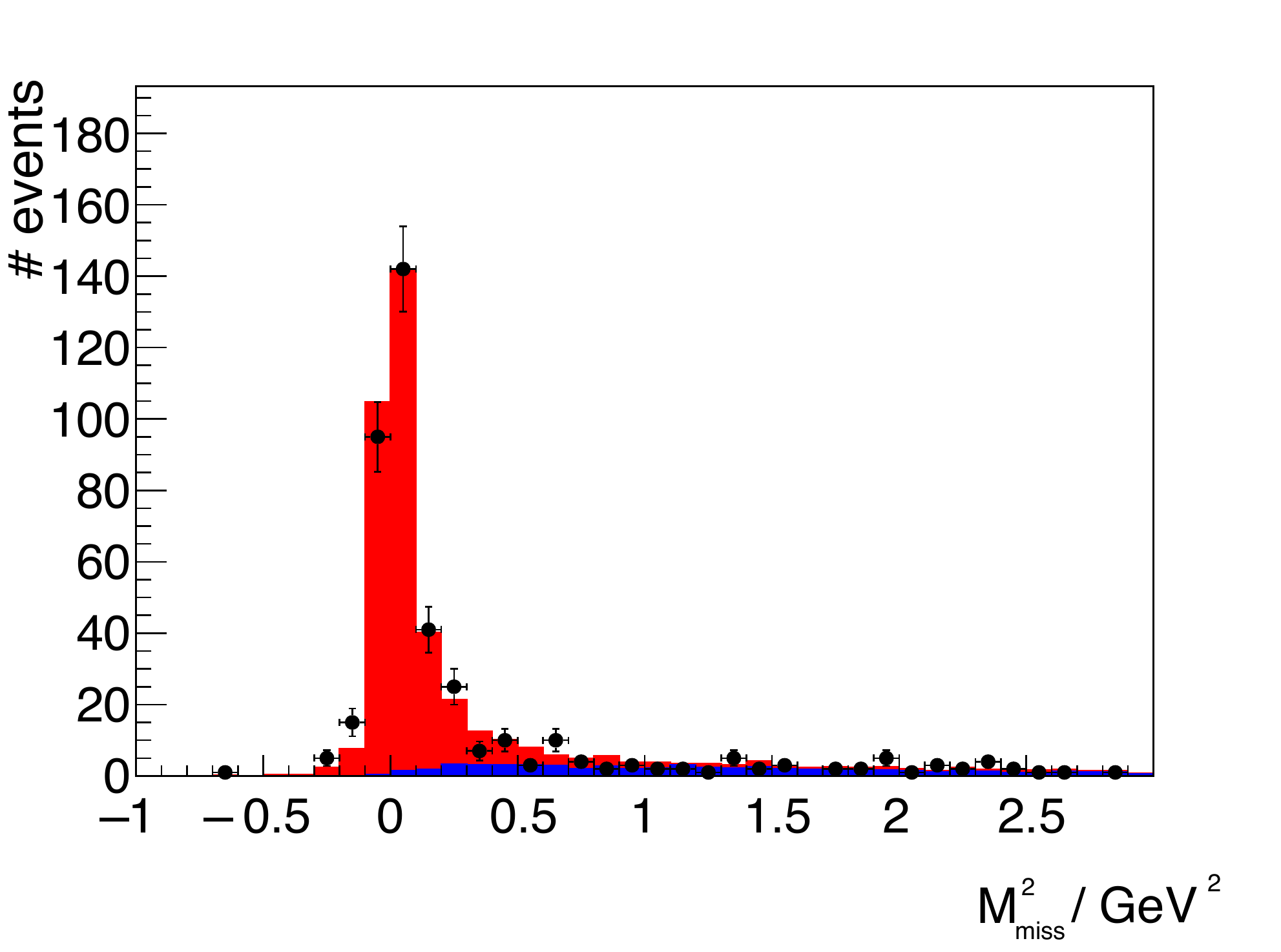}} \\
\subfigure[ \, $\chi \in [0, \pi/5)$ ]{\includegraphics[width=0.4\linewidth]{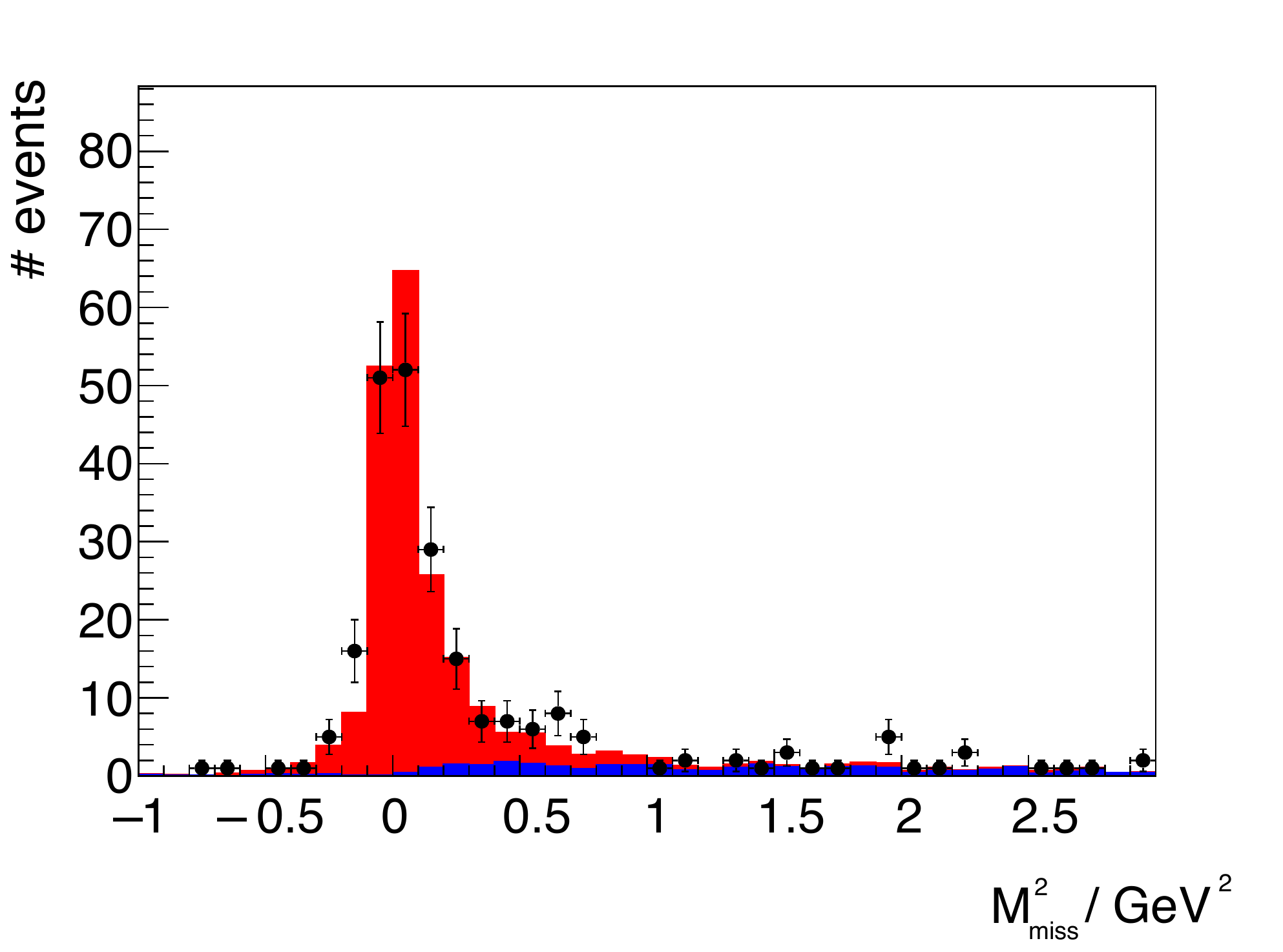}}
\caption{ The $M_{\rm miss}^2$ distributions after the likelihood fit for three representative bins in $w$, $\cos \theta_\ell$, and $\chi$ are shown. The PDFs were integrated over the corresponding bin boundaries for comparison between the data points and the signal and background contributions.}
\label{fig:B0Fitdiff}
\end{figure}

Each bin of the measured distributions of the hadronic recoil and angular variables is independently fitted for signal yields, and hence there is no assumption on the background distribution across these variables. 
The distributions are fitted in ten bins each using an equidistant binning (but extending the last bin in $w$ to account for the kinematic endpoint of the spectrum). This choice is a compromise of providing differential information, but also to reduce migration between the reconstructed and true underlying value of the kinematic quantities. A summary of the bin boundaries can be found in Table~\ref{tab:binning}. Figure~\ref{fig:B0Fitdiff} shows the $M_{\rm miss}^2$ distribution for three out of the forty differential bins for $w \in [1,1.05)$, $\cos \theta_\ell \in [0.8,1.0)$ and $\chi \in [0,\pi/5)$. The purity in each bin is very high and the unbinned PDFs have been integrated over the bins to allow for an easier comparison. The finite detector resolution and the mis-reconstruction of signal-side particles result in migration.
The inversion or unfolding of such effects for comparison to theory is discussed in Section~\ref{sec:unfolding}. 

The measured yields of the four kinematic variables are statistically correlated with each other as they a formed from the same reconstructed events. In order to simultaneously use information from $\{ w, \cos \theta_\ell, \cos \theta_v, \chi \}$ in the fit to determine $\left|V_{cb} \right|$, these correlations must be determined. This is achieved by using a bootstrapping procedure~\cite{bootstrap}: in each data subsample each data event is assigned a different Poisson weight $P(\nu = 1)$ and the yield extraction is repeated using these weighted events. A large number of subsamples is used to calculate the statistical correlation between the various bins. 

\begin{table*}
\begin{tabular}{c|c}
\hline\hline
 Variable & Bins \\ \hline
 $w$ & $[1.00, 1.05, 1.10, 1.15, 1.20, 1.25, 1.30, 1.35, 1.40, 1.45, 1.504]$ \\
 $\cos \theta_\ell$ & $[-1.0, -0.8, -0.6, -0.4, -0.2, 0.0, 0.2, 0.4, 0.6, 0.8, 1.0]$ \\
 $\cos \theta_v$ &$[-1.0, -0.8, -0.6, -0.4, -0.2, 0.0, 0.2, 0.4, 0.6, 0.8, 1.0]$ \\
 $\chi$ &$[0, \pi/5, 2 \pi/5, 3 \pi/5, 4 \pi/5, \pi, 6 \pi/5, 7 \pi/5, 8 \pi/5, 9\pi/5,2 \pi ]$ \\
\hline\hline
\end{tabular}
\caption{ The binning of the $w$, $\cos \theta_\ell$, $\cos \theta_v$, and $\chi$ distributions is shown.
\label{tab:binning}
 }
\end{table*}

\section{Unfolding of differential yields}\label{sec:unfolding}

Finite detector resolution and mis-reconstructed $D$ or $D^{*\, +}$-mesons result in migrations between the kinematic bins of $\{w, \cos \theta_\ell, \cos \theta_v, \chi \}$. Such migrations can be expressed in a detector response matrix of conditional probabilities, $\mathcal{P}( \text{reco bin} \, i \, | \, \text{true bin} \, j  )$, 
\begin{align}
 \mathcal{M}_{ij} & = \mathcal{P}( \text{reco bin} \, i \, | \, \text{true bin} \, j  ) \, ,
\end{align}
defined for each kinematic observable. The vector of extracted yields ${\bm \nu_{\rm sig} }$ for a given kinematic observable $x$ can then be related to the vector of differential branching fractions ${\bf \Delta \mathcal{B} / \Delta x }$ as
\begin{align}\label{eq:diffBF}
{\bf \Delta \mathcal{B} / \Delta x } & =  \left( {\bf \epsilon_{\rm reco} \epsilon_{\rm tag} } \right)^{-1} \times \mathcal{M}^{-1} \times {\bm \nu_{\rm sig}} \, \times \frac{1}{4 N_{B \bar B} \left( 1 + f_{+0} \right)^{-1} } \, .
\end{align}
Here the efficiency of reconstructing an event with a given true value of the kinematic variable $x$ inside a bin $j$ is parametrized as a diagonal matrix $ {\bf \epsilon_{\rm reco} \epsilon_{\rm tag} }$:
\begin{align}
\left( {\bf \epsilon_{\rm reco} \epsilon_{\rm tag} } \right)_{jj} & = \mathcal{A}( \text{true bin} \, j  ) \,  ,
\end{align}
which is often  called the acceptance $\mathcal{A}( \text{true bin} \, j  )$. Inverting the detector response in Eq.~\ref{eq:diffBF} is a non-trivial task: a direct numerical inversion of $\mathcal{M}$ leads to a large enhancement of statistical fluctuations. For the extraction of $\left| V_{cb} \right|$, the underlying theory is folded with the detector response and the acceptance. To preserve the measured spectra, the migration matrix is inverted using the SVD unfolding algorithm~\cite{Hocker:1995kb} 
Additional uncertainties are included in the error budget, introducing variations of $3 \sigma$ in the world average of the measured form factors to estimate the model error.

Table~\ref{tab:rates} lists the unfolded information converted in differential rates $\Delta \Gamma / \Delta x = \Delta \mathcal{B} / \Delta x  \times  \tau^{-1}$ using the $B^0$-lifetime of $\tau = 1.520 \, \text{ps}$. 
The full correlation matrix is provided in Appendix~\ref{app:correlation}.

\begin{table}[h!]\footnotesize
\begin{tabular}{c|ccc} 
\hline\hline 
 Variable & Bin & $\Delta \Gamma / \Delta x$ \,  [$10^{-15}\, $\text{GeV}] \\ \hline
 $w$ &1  & 	 $1.32 \pm 0.11$ \\
&2  & 	 $2.08 \pm 0.15$ \\
&3  & 	 $2.39 \pm 0.15$ \\
&4  & 	 $2.57 \pm 0.16$ \\
&5  & 	 $2.63 \pm 0.16$ \\
&6  & 	 $2.46 \pm 0.15$ \\
&7  & 	 $2.25 \pm 0.14$ \\
&8  & 	 $2.08 \pm 0.14$ \\
&9  & 	 $1.99 \pm 0.13$ \\
&10  & 	 $1.83 \pm 0.14$ \\
\hline
$\cos \theta_v$ &1  & 	 $2.80 \pm 0.20$ \\
&2  & 	 $2.30 \pm 0.14$ \\
&3  & 	 $1.95 \pm 0.13$ \\
&4  & 	 $1.70 \pm 0.12$ \\
&5  & 	 $1.58 \pm 0.12$ \\
&6  & 	 $1.65 \pm 0.11$ \\
&7  & 	 $1.77 \pm 0.12$ \\
&8  & 	 $2.00 \pm 0.14$ \\
&9  & 	 $2.50 \pm 0.17$ \\
&10  & 	 $3.19 \pm 0.25$ \\ \hline \hline
\end{tabular}
\hspace{5ex}
\begin{tabular}{c|ccc} 
\hline\hline 
 Variable & Bin & $\Delta \Gamma / \Delta x$ \,  [$10^{-15}\, $\text{GeV}] \\ \hline
$\cos \theta_\ell$ &1  & 	 $0.73 \pm 0.07$ \\
&2  & 	 $1.18 \pm 0.10$ \\
&3  & 	 $1.64 \pm 0.11$ \\
&4  & 	 $2.04 \pm 0.14$ \\
&5  & 	 $2.34 \pm 0.15$ \\
&6  & 	 $2.50 \pm 0.16$ \\
&7  & 	 $2.54 \pm 0.16$ \\
&8  & 	 $2.68 \pm 0.16$ \\
&9  & 	 $2.83 \pm 0.21$ \\
&10  & 	 $2.82 \pm 0.25$ \\ \hline
$\chi$ &1  & 	 $1.86 \pm 0.16$ \\
&2  & 	 $2.31 \pm 0.16$ \\
&3  & 	 $2.59 \pm 0.16$ \\
&4  & 	 $2.37 \pm 0.16$ \\
&5  & 	 $1.95 \pm 0.13$ \\
&6  & 	 $1.87 \pm 0.15$ \\
&7  & 	 $2.11 \pm 0.15$ \\
&8  & 	 $2.33 \pm 0.16$ \\
&9  & 	 $2.15 \pm 0.15$ \\
&10  & 	 $1.89 \pm 0.16$ \\
\hline\hline
\end{tabular}
\caption{ 
 The unfolded differential rates in units of $10^{-15} \, \text{GeV}$ are shown. 
 }
\label{tab:rates}
\end{table}

\section{Systematic uncertainties} \label{sec:syst}

There are several systematic uncertainties that affect the measured yields and branching fractions: Table~\ref{tab:syst_summary} summarizes the most important sources for the $\mathcal{B}(\bar B^0 \to D^{*\,+} \, \ell^- \, \bar \nu_\ell)$ branching fraction while the full set of systematics discussed in this section is also derived for the detector response and acceptance corrections, and propagated accordingly into the determination of $|V_{cb}|$ and the form factors. 

The largest systematic uncertainty on the branching fraction stems from the uncertainty on the tagging calibration, which is evaluated by shifting the central values of the correction factors, $C_{\rm tag}$, according to their corresponding statistical and correlated systematic uncertainties. The systematic uncertainties on the correction factors are due to the modelling of the $\bar B \to X \, \ell \, \bar \nu_\ell$ reference decay and the lepton PID efficiency errors and fake rates. Several replicas of the MC with these new correction factors are produced. The resulting differential spectra are almost unaffected by the change in tagging correction, thus only the impact on the overall acceptance is evaluated. The systematic error is estimated using a 68\% spread of the change in acceptance from many replicas and found to be of the order of $3.6$\%. The uncertainty on the tracking efficiency is 0.35\% per track and assumed to be fully correlated between all signal-side tracks. Possible differences on the tracking efficiency between simulated and measured events on the tagging side are absorbed in the tagging calibration factor. The uncertainty on the $\pi^0$ reconstruction efficiency is 2\%. Uncertainties on external parameters, such as the uncertainty on the number $B$-meson pairs ($N_{B \bar B}$) produced at Belle, the uncertainty on $f_{+0}$, and decay branching fractions are varied within their uncertainties and propagated to the final results. The constructed PDF shapes for signal and background components exhibit statistical uncertainties from the finite size of the MC samples. The resulting uncertainties are evaluated by bootstrapping the MC and replicas are produced by reweighing each MC event with a Poisson distribution of mean $\nu = 1$. For each MC replica the PDF shapes are rebuilt and the signal extraction on data is repeated. The resulting 68\% spread in the extracted yields are used as an estimator for the systematic uncertainty. The uncertainties from electron, muon, and kaon PID efficiency corrections are also evaluated  by producing replicas of the data: each replica is reweighed by a weight corresponding to the statistical and systematic error of the corresponding PID ratio, taking into account that the systematic errors are correlated over all events. This is done separately for each source and the 68\% spread on the final result is used as the uncertainty. For the construction of the systematic covariance matrix all uncertainties from a given single source are assumed to be fully correlated across all bins with the exception of the statistical uncertainty on the PDF shapes. 

\begin{table}[h!]
\vspace{2ex}
\begin{tabular}{l|c}
\hline\hline
Error Source & $\Delta \mathcal{B}$ [\%]  \\ \hline 
 Tagging Calibration & $3.6$  \\
 Tracking Efficiency & $1.6$\\
 $N_{B \bar B}$ & $1.4$  \\
 $f_{+0}$ & $1.1$  \\
 PDF shapes & $0.9$  \\
 $\pi^0$ Efficiency & $0.5$ \\
 $\mathcal{B}(D \to K \pi (\pi)(\pi))$   & $0.4$  \\
 $\mathcal{B}(D^* \to D \, \pi)$  & $0.2$  \\
 $\mathcal{B}(\bar B \to D^{**} \, \ell \, \bar\nu_\ell)$  & $0.2$  \\
 $e$ PID  & $0.2$  \\
 $\mu$ PID  & $0.1$  \\
 $\pi_{\rm slow}$ Eff.  & $0.1$ \\
 $\mathcal{B}(\bar B \to D \, \ell \, \bar\nu_\ell)$   & $<0.1$  \\
 $\bar B \to D^{(*,**)} \, \ell \, \bar \nu_\ell$ FFs  & $<0.1$  \\
 Lepton Fakerates & $< 0.1$  \\
 $K$ PID & $< 0.1$  \\
 \hline
 Total & 4.5 \\
\hline\hline
\end{tabular}
\caption{ Summary of the relative systematic errors ordered by importance in the total branching fraction measurement. }
\label{tab:syst_summary}
\end{table}

\begin{figure}
\includegraphics[width=0.48\textwidth,page=1,trim={0cm 0cm 0.5cm 0cm},clip]{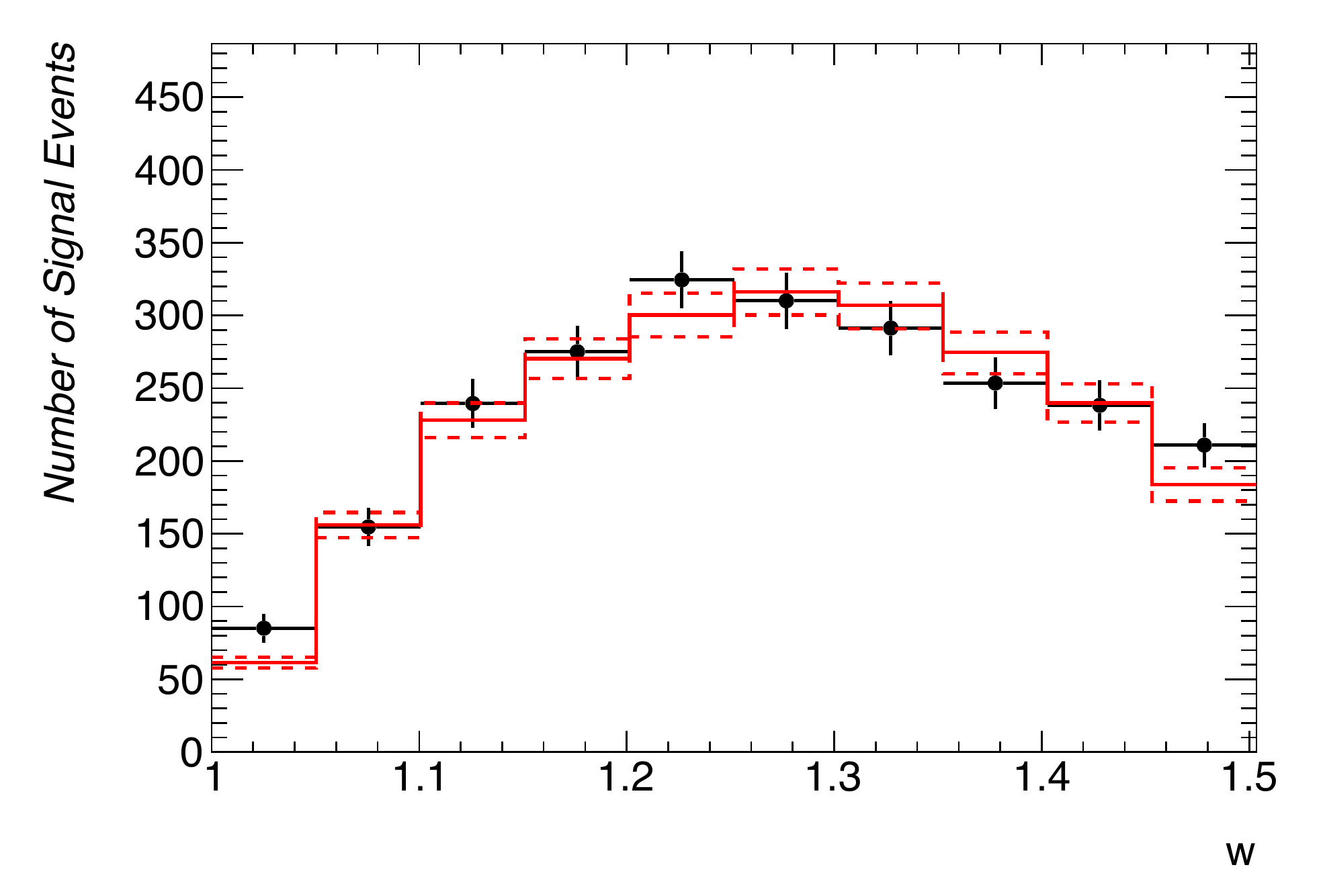} 
\includegraphics[width=0.48\textwidth,page=2,trim={0cm 0cm 0.5cm 0cm},clip]{./figures/B0_fitres.pdf} \\
\includegraphics[width=0.48\textwidth,page=3,trim={0cm 0cm 0.5cm 0cm},clip]{./figures/B0_fitres.pdf} 
\includegraphics[width=0.48\textwidth,page=4,trim={0cm 0cm 0.5cm 0cm},clip]{./figures/B0_fitres.pdf} \\
\caption{ 
 The fit result (solid red histograms) and the corresponding $\Delta \chi^2 + 1$ errors (dashed histograms) are shown. Details of the fit can be found in the text. 
}
\label{fig:FitRes}
\end{figure}

\section{Precise determination of $\left| V_{cb} \right|$}\label{sec:vcb}

The differential yields and their correlations are used to extract the form factor parameters defined in Section~\ref{sec:sltheory} and $\left| V_{cb} \right|$. This is done by constructing a $\chi^2$ function of the form
\begin{align}\label{eq:chi2}
 \chi^2 & = \left( {\bm \nu_{\rm sig}} - {\bm \nu_{\rm sig}^{\rm pred}}  \right) C^{-1} \, \left(  {\bm \nu_{\rm sig}} -  {\bm \nu_{\rm sig}^{\rm pred}} \right) + \chi_{\rm NP}^2 \, ,
\end{align}
with ${\bm \nu_{\rm sig} }$ the vector of measured yields, and \mbox{$ {\bm \nu_{\rm sig}^{\rm pred}} = \left(  {\bm \epsilon_{\rm reco} \epsilon_{\rm tag} } \right) \times \mathcal{M} \times  {\bm \Delta \Gamma / \Delta x } \, \tau $} the predicted number of signal events. The differential decay rate ${\bm \Delta \Gamma / \Delta x }$ is a function of the four parameters of interest, $\{ \left| V_{cb} \right|, \rho_{D^*}^2, R_1(1), R_2(1) \}$.

The covariance matrix $C$ contains all uncertainties associated to the signal extraction, while additional nuisance parameter terms $\chi_{\rm NP}^2$  are added to account for the uncertainties from multiplicative factors degenerate with $\left| V_{cb} \right|$.
The normalization of the universal form factor, $h_{A1}(1)$, is constrained to the lattice prediction of Ref.~\cite{Bailey:2014tva} (cf. Section~\ref{sec:sltheory}) using a constraint term of the form
\begin{align}
 \chi^2_{\rm la} & = \bigg(h_{A1}(1) - h_{A1}^{\rm la}(1) \bigg)^2 / \left( \sigma_{h_{A1}(1)}^{\rm la} \right)^2 \, ,
\end{align}
where $h_{A1}^{\rm la}(1) = 0.906$ and $\sigma_{h_{A1}(1)}^{\rm la} = 0.013$. Similar constraints are added to propagate the uncertainties from the full reconstruction algorithm calibration uncertainty, the error on the number of $B\bar B$-meson pairs, and the uncertainty on $f_{+0}$. 
\begin{figure}[t]
\includegraphics[width=0.48\textwidth,page=5,trim={0cm 0cm 0.5cm 0cm},clip]{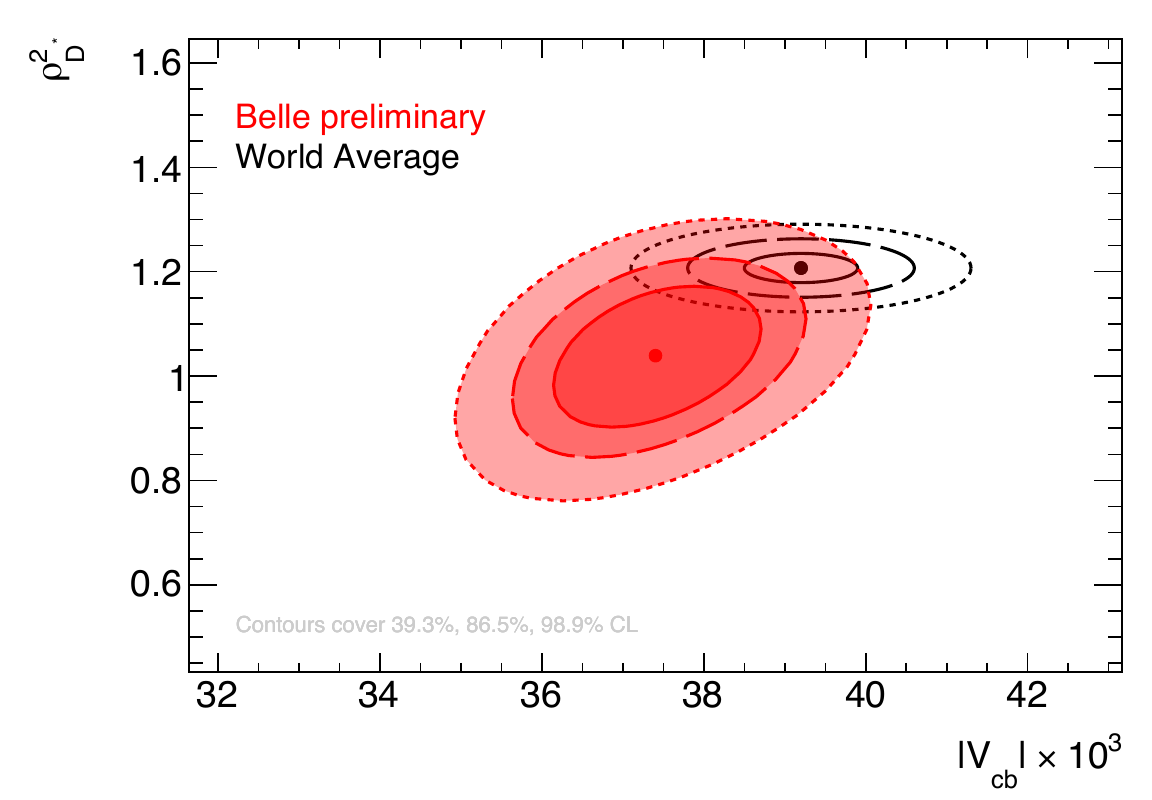} 
\includegraphics[width=0.48\textwidth,page=6,trim={0cm 0cm 0.5cm 0cm},clip]{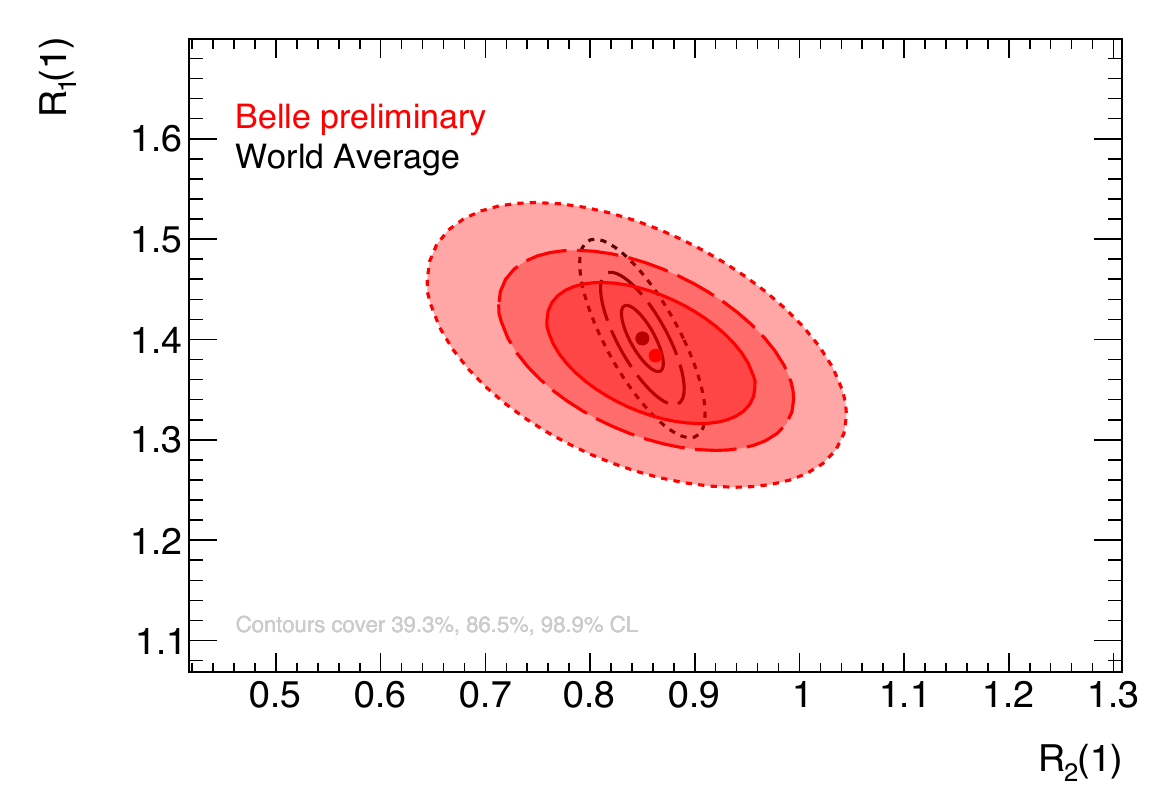} \\
\caption{ 
The best fit values for $\left| V_{cb} \right|$:$\rho_{D^*}^2$ and $R_1(1):R_2(1)$ with the corresponding $\Delta \chi^2 + 1$,  $\Delta \chi^2 + 2$, and $\Delta \chi^2 + 4$ contours are shown in red, while black contour shows the current world average from Ref.~\cite{hfag}.
}
\label{fig:FitRes2}
\end{figure}

Equation~\ref{eq:chi2} is numerically minimized to find the best fit values for $\left| V_{cb} \right|$ while the form factor parameters and their uncertainties are determined by scanning the $\Delta \chi^2 +1$ contours.  Figure~\ref{fig:FitRes} shows the fitted yields for all four variables as well as their respective best fit values and uncertainties. The fit has a $\chi^2 = 40.1$ with $40-4$ degrees of freedom, corresponding to a fit probability of $30\%$. We measure 
\begin{align}
 \left| V_{cb} \right| & =  \left( 37.4 \pm 1.3 \right) \times 10^{-3} \, ,
\end{align}
where the values of the form factors and of $\left| V_{cb} \right|$ are in good agreement with the current world average~\cite{hfag}. All numerical values are summarized in Table~\ref{tab:fitsummary}, and Figure~\ref{fig:FitRes2} shows the extracted values of $\left| V_{cb} \right| : \rho_{D^*}^2$ and $R_1(1): R_2(1)$. The correlation between $\{ \left| V_{cb} \right|, \rho_{D^*}^2, R_1(1), R_2(1) \}$ is determined to be
\begin{align}
 C = & \left( \begin{matrix}
    1 & 0.41 & -0.20 & -0.14 \\
    0.41 & 1 & 0.19 & -0.86 \\
    -0.20 & 0.19 & 1 & -0.46 \\    
    -0.14 & -0.86 & -0.46 & 1 \\     
 \end{matrix} \right) \, .
\end{align}
The results of the $\left| V_{cb} \right|$ and form factor fit to the unfolded differential branching fractions are provided in Appendix~\ref{app:VcbFit} .

\begin{table}[t]
\vspace{2ex}
\begin{tabular}{l|cc}
\hline\hline
Parameter & This result & World Average  \\ \hline 
$\left| V_{cb} \right| \times 10^3$ & $37.4 \pm 1.3$ & $39.2 \pm 0.7$ \\
$\rho_{D^*}^2$& $1.03 \pm 0.13$ & $1.21 \pm 0.03$ \\
$R_1(1)$& $1.38 \pm 0.07$ & $1.40 \pm 0.03$ \\
$R_2(1)$& $0.87 \pm 0.10$ & $0.85 \pm 0.02$ \\
\hline\hline
\end{tabular}

\caption{ The best-fit values of the fit is compared with the world average from Ref.~\cite{hfag}.  }
\label{tab:fitsummary}
\end{table}

\section{Summary and Conclusions}\label{sec:summary}

In this paper the precise determination of $\left| V_{cb} \right|$ using semileptonic $\bar B^0 \to D^{*\,+} \, \ell^- \, \bar \nu_\ell $ decays using a fully reconstructed dataset is reported. 
The total and differential signal yields in kinematic observables are extracted: the recoil parameter $w$ and three decay angles that fully characterize the $\bar B^0 \to D^{*\,+} \, \ell^- \, \bar \nu_\ell $ decay. The statistical correlations of the four variables are determined and the yields are unfolded as binned differential decay widths.
 From the total yield the  $\bar B^0 \to D^{*\,+} \, \ell^- \, \bar \nu_\ell$ branching fraction is determined to be 
 \begin{align}
 \mathcal{B}(\bar B^0 \to D^{*\,+} \, \ell^- \, \bar \nu_\ell) & = \left(4.95 \pm 0.11 \pm 0.22 \right)\times 10^{-2} \, ,
\end{align}
which is in good agreement with the current world average of Ref.~\cite{pdg}. The value of $\left| V_{cb} \right|$ is determined by simultaneously fitting all four kinematic variables: 
\begin{align}
 \left| V_{cb} \right| & = \left( 37.4 \pm 1.3 \right) \times 10^{-3} \, ,
\end{align}
 which is in good agreement with the current world average ~\cite{hfag}. The unfolded differential decay rates are reported for the first time, which can be directly compared to theoretical expectations.
Finally, using the full correlation matrix of the extracted form factor parameters, a prediction for the ratio of semileptonic decays with $\tau$ and light lepton final states can be computed,
\begin{align}
 R(D^*) = \frac{ \mathcal{B}( \bar B \to D^{*} \, \tau \,  \bar \nu_\tau)  }{\mathcal{B}( \bar B \to D^{*} \, \ell \,  \bar \nu_\ell)} \, ,
\end{align}
with $\ell = e$ or $\mu$. This is of interest as many recent measurements report a significant enhancement over the SM expectation of this ratio. Using the fitted values of $\rho_{D^*}^2$, $R_1(1)$ and $R_2(1)$ and the associated uncertainties we obtain
\begin{align}
 R(D^*)_{\rm SM} & = 0.242 \pm 0.005 \, ,
\end{align}
based on a value of $R_0(1) = 1.14 \pm 0.11$ from Ref.~\cite{Fajfer:2012vx} for the form factor ratio unconstrained by light lepton measurements. This ratio is slightly lower than the prediction from Ref.~\cite{Fajfer:2012vx} of $ R(D^*)_{\rm SM} = 0.252 \pm 0.003$ and in tension with the current world average~\cite{hfag}
\begin{align}
 R(D^*)_{\rm wa} & = 0.310 \pm 0.015 \pm 0.008 \, ,
\end{align}
where the first error is statistical and the second from systematic uncertainties. The tension between the predicted and the observed values is approximately $3.8$ standard deviations.

\vspace{0.3cm}
\section*{Acknowledgments}
We thank Stefan Schacht and Andrew Kobach to point out an inconsistency in the $\left| V_{cb} \right|$ fit results.
We thank the KEKB group for the excellent operation of the
accelerator; the KEK cryogenics group for the efficient
operation of the solenoid; and the KEK computer group,
the National Institute of Informatics, and the 
PNNL/EMSL computing group for valuable computing
and SINET5 network support.  We acknowledge support from
the Ministry of Education, Culture, Sports, Science, and
Technology (MEXT) of Japan, the Japan Society for the 
Promotion of Science (JSPS), and the Tau-Lepton Physics 
Research Center of Nagoya University; 
the Australian Research Council;
Austrian Science Fund under Grant No.~P 26794-N20;
the National Natural Science Foundation of China under Contracts 
No.~10575109, No.~10775142, No.~10875115, No.~11175187, No.~11475187, 
No.~11521505 and No.~11575017;
the Chinese Academy of Science Center for Excellence in Particle Physics; 
the Ministry of Education, Youth and Sports of the Czech
Republic under Contract No.~LG14034;
the Carl Zeiss Foundation, the Deutsche Forschungsgemeinschaft, the
Excellence Cluster Universe, and the VolkswagenStiftung;
the Department of Science and Technology of India; 
the Istituto Nazionale di Fisica Nucleare of Italy; 
the WCU program of the Ministry of Education, National Research Foundation (NRF) 
of Korea Grants No.~2011-0029457,  No.~2012-0008143,  
No.~2014R1A2A2A01005286, 
No.~2014R1A2A2A01002734, No.~2015R1A2A2A01003280,
No.~2015H1A2A1033649, No.~2016R1D1A1B01010135, No.~2016K1A3A7A09005603, No.~2016K1A3A7A09005604, No.~2016R1D1A1B02012900,
No.~2016K1A3A7A09005606, No.~NRF-2013K1A3A7A06056592; 
the Brain Korea 21-Plus program and Radiation Science Research Institute;
the Polish Ministry of Science and Higher Education and 
the National Science Center;
the Ministry of Education and Science of the Russian Federation and
the Russian Foundation for Basic Research;
the Slovenian Research Agency;
Ikerbasque, Basque Foundation for Science and
the Euskal Herriko Unibertsitatea (UPV/EHU) under program UFI 11/55 (Spain);
the Swiss National Science Foundation; 
the Ministry of Education and the Ministry of Science and Technology of Taiwan;
and the U.S.\ Department of Energy and the National Science Foundation.

\begin{appendix}

\section{Correlation matrix of the unfolded spectra}\label{app:correlation}

The correlation matrix of the unfolded differential rates is listed below: The full error covariance can be obtained by combining the quoted error in Table~\ref{tab:rates} with these values. The ordering of the correlations is $\{w, \cos \theta_v, \cos \theta_\ell, \chi \}$.

\begin{sidewaystable} {\tiny 
\vspace{30ex}
\resizebox{\columnwidth}{!}{
\begin{tabular}{c|lllllllllllllllllllllllllllllllllllllllllllllllllllllllllllllllllllllllllllllllllllllllllllllllllllllllllllllllll}  
Bin & 1 & 2 & 3 & 4 & 5 & 6 & 7 & 8 & 9 & 10 & 11 & 12 & 13 & 14 & 15 & 16 & 17 & 18 & 19 & 20 & 21 & 22 & 23 & 24 & 25 & 26 & 27 & 28 & 29 & 30 & 31 & 32 & 33 & 34 & 35 & 36 & 37 & 38 & 39 & 40 \\  \hline
1 & 1.00 & 0.92 & 0.60 & 0.35 & 0.30 & 0.34 & 0.38 & 0.40 & 0.37 & 0.31 & 0.32 & 0.33 & 0.39 & 0.42 & 0.43 & 0.42 & 0.40 & 0.39 & 0.38 & 0.35 & 0.31 & 0.35 & 0.39 & 0.40 & 0.41 & 0.41 & 0.38 & 0.38 & 0.37 & 0.33 & 0.28 & 0.35 & 0.40 & 0.41 & 0.41 & 0.37 & 0.40 & 0.42 & 0.42 & 0.35\\
2 & 0.92 & 1.00 & 0.80 & 0.50 & 0.39 & 0.40 & 0.44 & 0.45 & 0.43 & 0.36 & 0.35 & 0.38 & 0.42 & 0.46 & 0.46 & 0.46 & 0.46 & 0.45 & 0.44 & 0.38 & 0.34 & 0.41 & 0.46 & 0.46 & 0.47 & 0.46 & 0.44 & 0.43 & 0.39 & 0.34 & 0.32 & 0.39 & 0.46 & 0.46 & 0.46 & 0.38 & 0.44 & 0.46 & 0.45 & 0.38\\
3 & 0.60 & 0.80 & 1.00 & 0.83 & 0.60 & 0.52 & 0.52 & 0.55 & 0.52 & 0.46 & 0.38 & 0.43 & 0.53 & 0.54 & 0.55 & 0.53 & 0.52 & 0.55 & 0.52 & 0.46 & 0.40 & 0.47 & 0.54 & 0.57 & 0.58 & 0.56 & 0.50 & 0.50 & 0.52 & 0.47 & 0.40 & 0.47 & 0.55 & 0.56 & 0.55 & 0.52 & 0.55 & 0.56 & 0.53 & 0.47\\
4 & 0.35 & 0.50 & 0.83 & 1.00 & 0.83 & 0.65 & 0.55 & 0.57 & 0.54 & 0.50 & 0.37 & 0.44 & 0.59 & 0.59 & 0.59 & 0.55 & 0.53 & 0.57 & 0.55 & 0.50 & 0.40 & 0.48 & 0.55 & 0.59 & 0.60 & 0.59 & 0.52 & 0.54 & 0.57 & 0.52 & 0.43 & 0.48 & 0.56 & 0.59 & 0.59 & 0.59 & 0.58 & 0.58 & 0.56 & 0.52\\
5 & 0.30 & 0.39 & 0.60 & 0.83 & 1.00 & 0.85 & 0.59 & 0.49 & 0.50 & 0.45 & 0.36 & 0.45 & 0.53 & 0.56 & 0.53 & 0.53 & 0.52 & 0.50 & 0.52 & 0.47 & 0.37 & 0.45 & 0.52 & 0.55 & 0.56 & 0.55 & 0.54 & 0.57 & 0.47 & 0.39 & 0.37 & 0.43 & 0.53 & 0.54 & 0.55 & 0.46 & 0.52 & 0.52 & 0.54 & 0.48\\
6 & 0.34 & 0.40 & 0.52 & 0.65 & 0.85 & 1.00 & 0.81 & 0.60 & 0.49 & 0.42 & 0.37 & 0.45 & 0.57 & 0.58 & 0.56 & 0.55 & 0.52 & 0.52 & 0.52 & 0.47 & 0.39 & 0.47 & 0.54 & 0.57 & 0.58 & 0.57 & 0.56 & 0.57 & 0.50 & 0.43 & 0.41 & 0.47 & 0.54 & 0.55 & 0.56 & 0.51 & 0.54 & 0.55 & 0.55 & 0.48\\
7 & 0.38 & 0.44 & 0.52 & 0.55 & 0.59 & 0.81 & 1.00 & 0.85 & 0.61 & 0.46 & 0.37 & 0.45 & 0.58 & 0.59 & 0.59 & 0.56 & 0.53 & 0.55 & 0.54 & 0.49 & 0.41 & 0.48 & 0.55 & 0.56 & 0.57 & 0.57 & 0.56 & 0.55 & 0.53 & 0.46 & 0.44 & 0.49 & 0.56 & 0.58 & 0.61 & 0.58 & 0.57 & 0.56 & 0.53 & 0.46\\
8 & 0.40 & 0.45 & 0.55 & 0.57 & 0.49 & 0.60 & 0.85 & 1.00 & 0.83 & 0.63 & 0.39 & 0.41 & 0.57 & 0.58 & 0.63 & 0.59 & 0.54 & 0.58 & 0.56 & 0.52 & 0.40 & 0.46 & 0.53 & 0.57 & 0.57 & 0.58 & 0.55 & 0.55 & 0.58 & 0.54 & 0.45 & 0.51 & 0.55 & 0.60 & 0.61 & 0.63 & 0.58 & 0.59 & 0.57 & 0.51\\
9 & 0.37 & 0.43 & 0.52 & 0.54 & 0.50 & 0.49 & 0.61 & 0.83 & 1.00 & 0.91 & 0.38 & 0.41 & 0.51 & 0.53 & 0.58 & 0.57 & 0.54 & 0.54 & 0.53 & 0.50 & 0.36 & 0.42 & 0.51 & 0.55 & 0.55 & 0.55 & 0.54 & 0.54 & 0.52 & 0.47 & 0.42 & 0.50 & 0.53 & 0.54 & 0.55 & 0.53 & 0.53 & 0.54 & 0.54 & 0.49\\
10 & 0.31 & 0.36 & 0.46 & 0.50 & 0.45 & 0.42 & 0.46 & 0.63 & 0.91 & 1.00 & 0.33 & 0.36 & 0.46 & 0.48 & 0.53 & 0.51 & 0.48 & 0.50 & 0.48 & 0.47 & 0.32 & 0.38 & 0.46 & 0.50 & 0.50 & 0.49 & 0.48 & 0.49 & 0.50 & 0.46 & 0.39 & 0.47 & 0.48 & 0.47 & 0.47 & 0.49 & 0.47 & 0.48 & 0.48 & 0.46\\
11 & 0.32 & 0.35 & 0.38 & 0.37 & 0.36 & 0.37 & 0.37 & 0.39 & 0.38 & 0.33 & 1.00 & 0.66 & 0.36 & 0.29 & 0.33 & 0.33 & 0.32 & 0.34 & 0.34 & 0.31 & 0.31 & 0.35 & 0.39 & 0.39 & 0.40 & 0.39 & 0.38 & 0.37 & 0.39 & 0.33 & 0.28 & 0.37 & 0.42 & 0.41 & 0.41 & 0.39 & 0.41 & 0.39 & 0.38 & 0.35\\
12 & 0.33 & 0.38 & 0.43 & 0.44 & 0.45 & 0.45 & 0.45 & 0.41 & 0.41 & 0.36 & 0.66 & 1.00 & 0.75 & 0.53 & 0.39 & 0.36 & 0.36 & 0.37 & 0.39 & 0.37 & 0.37 & 0.42 & 0.47 & 0.45 & 0.46 & 0.45 & 0.43 & 0.43 & 0.39 & 0.33 & 0.33 & 0.40 & 0.47 & 0.47 & 0.49 & 0.40 & 0.44 & 0.44 & 0.42 & 0.38\\
13 & 0.39 & 0.42 & 0.53 & 0.59 & 0.53 & 0.57 & 0.58 & 0.57 & 0.51 & 0.46 & 0.36 & 0.75 & 1.00 & 0.87 & 0.69 & 0.54 & 0.45 & 0.52 & 0.51 & 0.48 & 0.43 & 0.50 & 0.56 & 0.59 & 0.59 & 0.58 & 0.52 & 0.56 & 0.60 & 0.54 & 0.44 & 0.50 & 0.56 & 0.62 & 0.63 & 0.63 & 0.58 & 0.58 & 0.56 & 0.53\\
14 & 0.42 & 0.46 & 0.54 & 0.59 & 0.56 & 0.58 & 0.59 & 0.58 & 0.53 & 0.48 & 0.29 & 0.53 & 0.87 & 1.00 & 0.89 & 0.71 & 0.55 & 0.51 & 0.48 & 0.45 & 0.42 & 0.50 & 0.58 & 0.60 & 0.61 & 0.59 & 0.54 & 0.56 & 0.57 & 0.51 & 0.43 & 0.48 & 0.57 & 0.62 & 0.63 & 0.60 & 0.57 & 0.57 & 0.57 & 0.53\\
15 & 0.43 & 0.46 & 0.55 & 0.59 & 0.53 & 0.56 & 0.59 & 0.63 & 0.58 & 0.53 & 0.33 & 0.39 & 0.69 & 0.89 & 1.00 & 0.90 & 0.69 & 0.59 & 0.47 & 0.44 & 0.39 & 0.46 & 0.56 & 0.62 & 0.63 & 0.61 & 0.55 & 0.58 & 0.63 & 0.58 & 0.45 & 0.52 & 0.58 & 0.64 & 0.64 & 0.66 & 0.60 & 0.60 & 0.60 & 0.56\\
16 & 0.42 & 0.46 & 0.53 & 0.55 & 0.53 & 0.55 & 0.56 & 0.59 & 0.57 & 0.51 & 0.33 & 0.36 & 0.54 & 0.71 & 0.90 & 1.00 & 0.87 & 0.66 & 0.47 & 0.38 & 0.39 & 0.46 & 0.55 & 0.60 & 0.61 & 0.60 & 0.56 & 0.59 & 0.56 & 0.50 & 0.41 & 0.49 & 0.56 & 0.60 & 0.60 & 0.58 & 0.58 & 0.60 & 0.60 & 0.53\\
17 & 0.40 & 0.46 & 0.52 & 0.53 & 0.52 & 0.52 & 0.53 & 0.54 & 0.54 & 0.48 & 0.32 & 0.36 & 0.45 & 0.55 & 0.69 & 0.87 & 1.00 & 0.85 & 0.59 & 0.37 & 0.40 & 0.46 & 0.52 & 0.55 & 0.58 & 0.58 & 0.55 & 0.57 & 0.49 & 0.43 & 0.38 & 0.46 & 0.53 & 0.54 & 0.55 & 0.50 & 0.55 & 0.58 & 0.58 & 0.50\\
18 & 0.39 & 0.45 & 0.55 & 0.57 & 0.50 & 0.52 & 0.55 & 0.58 & 0.54 & 0.50 & 0.34 & 0.37 & 0.52 & 0.51 & 0.59 & 0.66 & 0.85 & 1.00 & 0.83 & 0.57 & 0.42 & 0.48 & 0.53 & 0.57 & 0.59 & 0.60 & 0.54 & 0.56 & 0.59 & 0.54 & 0.44 & 0.50 & 0.55 & 0.59 & 0.59 & 0.62 & 0.61 & 0.61 & 0.60 & 0.55\\
19 & 0.38 & 0.44 & 0.52 & 0.55 & 0.52 & 0.52 & 0.54 & 0.56 & 0.53 & 0.48 & 0.34 & 0.39 & 0.51 & 0.48 & 0.47 & 0.47 & 0.59 & 0.83 & 1.00 & 0.85 & 0.39 & 0.45 & 0.51 & 0.53 & 0.56 & 0.57 & 0.52 & 0.52 & 0.52 & 0.47 & 0.44 & 0.49 & 0.52 & 0.56 & 0.57 & 0.55 & 0.57 & 0.57 & 0.56 & 0.50\\
20 & 0.35 & 0.38 & 0.46 & 0.50 & 0.47 & 0.47 & 0.49 & 0.52 & 0.50 & 0.47 & 0.31 & 0.37 & 0.48 & 0.45 & 0.44 & 0.38 & 0.37 & 0.57 & 0.85 & 1.00 & 0.32 & 0.38 & 0.45 & 0.50 & 0.52 & 0.52 & 0.45 & 0.45 & 0.51 & 0.48 & 0.41 & 0.47 & 0.47 & 0.50 & 0.50 & 0.54 & 0.53 & 0.53 & 0.52 & 0.47\\
21 & 0.31 & 0.34 & 0.40 & 0.40 & 0.37 & 0.39 & 0.41 & 0.40 & 0.36 & 0.32 & 0.31 & 0.37 & 0.43 & 0.42 & 0.39 & 0.39 & 0.40 & 0.42 & 0.39 & 0.32 & 1.00 & 0.93 & 0.67 & 0.36 & 0.28 & 0.31 & 0.33 & 0.37 & 0.34 & 0.29 & 0.29 & 0.33 & 0.36 & 0.39 & 0.43 & 0.41 & 0.43 & 0.40 & 0.38 & 0.33\\
22 & 0.35 & 0.41 & 0.47 & 0.48 & 0.45 & 0.47 & 0.48 & 0.46 & 0.42 & 0.38 & 0.35 & 0.42 & 0.50 & 0.50 & 0.46 & 0.46 & 0.46 & 0.48 & 0.45 & 0.38 & 0.93 & 1.00 & 0.85 & 0.53 & 0.38 & 0.37 & 0.40 & 0.43 & 0.39 & 0.33 & 0.34 & 0.39 & 0.44 & 0.46 & 0.50 & 0.46 & 0.48 & 0.45 & 0.44 & 0.39\\
23 & 0.39 & 0.46 & 0.54 & 0.55 & 0.52 & 0.54 & 0.55 & 0.53 & 0.51 & 0.46 & 0.39 & 0.47 & 0.56 & 0.58 & 0.56 & 0.55 & 0.52 & 0.53 & 0.51 & 0.45 & 0.67 & 0.85 & 1.00 & 0.83 & 0.60 & 0.47 & 0.46 & 0.49 & 0.47 & 0.41 & 0.38 & 0.44 & 0.51 & 0.54 & 0.57 & 0.53 & 0.55 & 0.53 & 0.51 & 0.47\\
24 & 0.40 & 0.46 & 0.57 & 0.59 & 0.55 & 0.57 & 0.56 & 0.57 & 0.55 & 0.50 & 0.39 & 0.45 & 0.59 & 0.60 & 0.62 & 0.60 & 0.55 & 0.57 & 0.53 & 0.50 & 0.36 & 0.53 & 0.83 & 1.00 & 0.86 & 0.63 & 0.49 & 0.50 & 0.54 & 0.50 & 0.41 & 0.48 & 0.54 & 0.58 & 0.60 & 0.59 & 0.59 & 0.58 & 0.55 & 0.50\\
25 & 0.41 & 0.47 & 0.58 & 0.60 & 0.56 & 0.58 & 0.57 & 0.57 & 0.55 & 0.50 & 0.40 & 0.46 & 0.59 & 0.61 & 0.63 & 0.61 & 0.58 & 0.59 & 0.56 & 0.52 & 0.28 & 0.38 & 0.60 & 0.86 & 1.00 & 0.86 & 0.61 & 0.51 & 0.52 & 0.50 & 0.41 & 0.48 & 0.55 & 0.59 & 0.60 & 0.58 & 0.59 & 0.59 & 0.57 & 0.50\\
26 & 0.41 & 0.46 & 0.56 & 0.59 & 0.55 & 0.57 & 0.57 & 0.58 & 0.55 & 0.49 & 0.39 & 0.45 & 0.58 & 0.59 & 0.61 & 0.60 & 0.58 & 0.60 & 0.57 & 0.52 & 0.31 & 0.37 & 0.47 & 0.63 & 0.86 & 1.00 & 0.82 & 0.60 & 0.53 & 0.49 & 0.43 & 0.50 & 0.54 & 0.59 & 0.60 & 0.58 & 0.58 & 0.59 & 0.57 & 0.50\\
27 & 0.38 & 0.44 & 0.50 & 0.52 & 0.54 & 0.56 & 0.56 & 0.55 & 0.54 & 0.48 & 0.38 & 0.43 & 0.52 & 0.54 & 0.55 & 0.56 & 0.55 & 0.54 & 0.52 & 0.45 & 0.33 & 0.40 & 0.46 & 0.49 & 0.61 & 0.82 & 1.00 & 0.83 & 0.53 & 0.38 & 0.39 & 0.47 & 0.53 & 0.54 & 0.56 & 0.51 & 0.53 & 0.53 & 0.53 & 0.46\\
28 & 0.38 & 0.43 & 0.50 & 0.54 & 0.57 & 0.57 & 0.55 & 0.55 & 0.54 & 0.49 & 0.37 & 0.43 & 0.56 & 0.56 & 0.58 & 0.59 & 0.57 & 0.56 & 0.52 & 0.45 & 0.37 & 0.43 & 0.49 & 0.50 & 0.51 & 0.60 & 0.83 & 1.00 & 0.74 & 0.46 & 0.40 & 0.49 & 0.55 & 0.55 & 0.56 & 0.55 & 0.57 & 0.55 & 0.55 & 0.49\\
29 & 0.37 & 0.39 & 0.52 & 0.57 & 0.47 & 0.50 & 0.53 & 0.58 & 0.52 & 0.50 & 0.39 & 0.39 & 0.60 & 0.57 & 0.63 & 0.56 & 0.49 & 0.59 & 0.52 & 0.51 & 0.34 & 0.39 & 0.47 & 0.54 & 0.52 & 0.53 & 0.53 & 0.74 & 1.00 & 0.87 & 0.44 & 0.51 & 0.54 & 0.59 & 0.58 & 0.70 & 0.61 & 0.59 & 0.56 & 0.54\\
30 & 0.33 & 0.34 & 0.47 & 0.52 & 0.39 & 0.43 & 0.46 & 0.54 & 0.47 & 0.46 & 0.33 & 0.33 & 0.54 & 0.51 & 0.58 & 0.50 & 0.43 & 0.54 & 0.47 & 0.48 & 0.29 & 0.33 & 0.41 & 0.50 & 0.50 & 0.49 & 0.38 & 0.46 & 0.87 & 1.00 & 0.40 & 0.46 & 0.47 & 0.55 & 0.53 & 0.66 & 0.55 & 0.55 & 0.52 & 0.52\\
31 & 0.28 & 0.32 & 0.40 & 0.43 & 0.37 & 0.41 & 0.44 & 0.45 & 0.42 & 0.39 & 0.28 & 0.33 & 0.44 & 0.43 & 0.45 & 0.41 & 0.38 & 0.44 & 0.44 & 0.41 & 0.29 & 0.34 & 0.38 & 0.41 & 0.41 & 0.43 & 0.39 & 0.40 & 0.44 & 0.40 & 1.00 & 0.86 & 0.48 & 0.37 & 0.39 & 0.45 & 0.44 & 0.38 & 0.29 & 0.26\\
32 & 0.35 & 0.39 & 0.47 & 0.48 & 0.43 & 0.47 & 0.49 & 0.51 & 0.50 & 0.47 & 0.37 & 0.40 & 0.50 & 0.48 & 0.52 & 0.49 & 0.46 & 0.50 & 0.49 & 0.47 & 0.33 & 0.39 & 0.44 & 0.48 & 0.48 & 0.50 & 0.47 & 0.49 & 0.51 & 0.46 & 0.86 & 1.00 & 0.77 & 0.55 & 0.47 & 0.49 & 0.49 & 0.45 & 0.37 & 0.32\\
33 & 0.40 & 0.46 & 0.55 & 0.56 & 0.53 & 0.54 & 0.56 & 0.55 & 0.53 & 0.48 & 0.42 & 0.47 & 0.56 & 0.57 & 0.58 & 0.56 & 0.53 & 0.55 & 0.52 & 0.47 & 0.36 & 0.44 & 0.51 & 0.54 & 0.55 & 0.54 & 0.53 & 0.55 & 0.54 & 0.47 & 0.48 & 0.77 & 1.00 & 0.85 & 0.65 & 0.52 & 0.51 & 0.49 & 0.48 & 0.43\\
34 & 0.41 & 0.46 & 0.56 & 0.59 & 0.54 & 0.55 & 0.58 & 0.60 & 0.54 & 0.47 & 0.41 & 0.47 & 0.62 & 0.62 & 0.64 & 0.60 & 0.54 & 0.59 & 0.56 & 0.50 & 0.39 & 0.46 & 0.54 & 0.58 & 0.59 & 0.59 & 0.54 & 0.55 & 0.59 & 0.55 & 0.37 & 0.55 & 0.85 & 1.00 & 0.87 & 0.67 & 0.55 & 0.52 & 0.53 & 0.50\\
35 & 0.41 & 0.46 & 0.55 & 0.59 & 0.55 & 0.56 & 0.61 & 0.61 & 0.55 & 0.47 & 0.41 & 0.49 & 0.63 & 0.63 & 0.64 & 0.60 & 0.55 & 0.59 & 0.57 & 0.50 & 0.43 & 0.50 & 0.57 & 0.60 & 0.60 & 0.60 & 0.56 & 0.56 & 0.58 & 0.53 & 0.39 & 0.47 & 0.65 & 0.87 & 1.00 & 0.83 & 0.67 & 0.54 & 0.50 & 0.47\\
36 & 0.37 & 0.38 & 0.52 & 0.59 & 0.46 & 0.51 & 0.58 & 0.63 & 0.53 & 0.49 & 0.39 & 0.40 & 0.63 & 0.60 & 0.66 & 0.58 & 0.50 & 0.62 & 0.55 & 0.54 & 0.41 & 0.46 & 0.53 & 0.59 & 0.58 & 0.58 & 0.51 & 0.55 & 0.70 & 0.66 & 0.45 & 0.49 & 0.52 & 0.67 & 0.83 & 1.00 & 0.85 & 0.66 & 0.51 & 0.47\\
37 & 0.40 & 0.44 & 0.55 & 0.58 & 0.52 & 0.54 & 0.57 & 0.58 & 0.53 & 0.47 & 0.41 & 0.44 & 0.58 & 0.57 & 0.60 & 0.58 & 0.55 & 0.61 & 0.57 & 0.53 & 0.43 & 0.48 & 0.55 & 0.59 & 0.59 & 0.58 & 0.53 & 0.57 & 0.61 & 0.55 & 0.44 & 0.49 & 0.51 & 0.55 & 0.67 & 0.85 & 1.00 & 0.85 & 0.57 & 0.42\\
38 & 0.42 & 0.46 & 0.56 & 0.58 & 0.52 & 0.55 & 0.56 & 0.59 & 0.54 & 0.48 & 0.39 & 0.44 & 0.58 & 0.57 & 0.60 & 0.60 & 0.58 & 0.61 & 0.57 & 0.53 & 0.40 & 0.45 & 0.53 & 0.58 & 0.59 & 0.59 & 0.53 & 0.55 & 0.59 & 0.55 & 0.38 & 0.45 & 0.49 & 0.52 & 0.54 & 0.66 & 0.85 & 1.00 & 0.81 & 0.57\\
39 & 0.42 & 0.45 & 0.53 & 0.56 & 0.54 & 0.55 & 0.53 & 0.57 & 0.54 & 0.48 & 0.38 & 0.42 & 0.56 & 0.57 & 0.60 & 0.60 & 0.58 & 0.60 & 0.56 & 0.52 & 0.38 & 0.44 & 0.51 & 0.55 & 0.57 & 0.57 & 0.53 & 0.55 & 0.56 & 0.52 & 0.29 & 0.37 & 0.48 & 0.53 & 0.50 & 0.51 & 0.57 & 0.81 & 1.00 & 0.88\\
40 & 0.35 & 0.38 & 0.47 & 0.52 & 0.48 & 0.48 & 0.46 & 0.51 & 0.49 & 0.46 & 0.35 & 0.38 & 0.53 & 0.53 & 0.56 & 0.53 & 0.50 & 0.55 & 0.50 & 0.47 & 0.33 & 0.39 & 0.47 & 0.50 & 0.50 & 0.50 & 0.46 & 0.49 & 0.54 & 0.52 & 0.26 & 0.32 & 0.43 & 0.50 & 0.47 & 0.47 & 0.42 & 0.57 & 0.88 & 1.00\\
\end{tabular} 
}}
\end{sidewaystable}

\clearpage

\section{$\left| V_{cb} \right|$ fit of the unfolded spectra}\label{app:VcbFit}

\begin{figure}[h]
\includegraphics[width=0.48\textwidth,page=5,trim={0cm 0cm 0.5cm 0cm},clip]{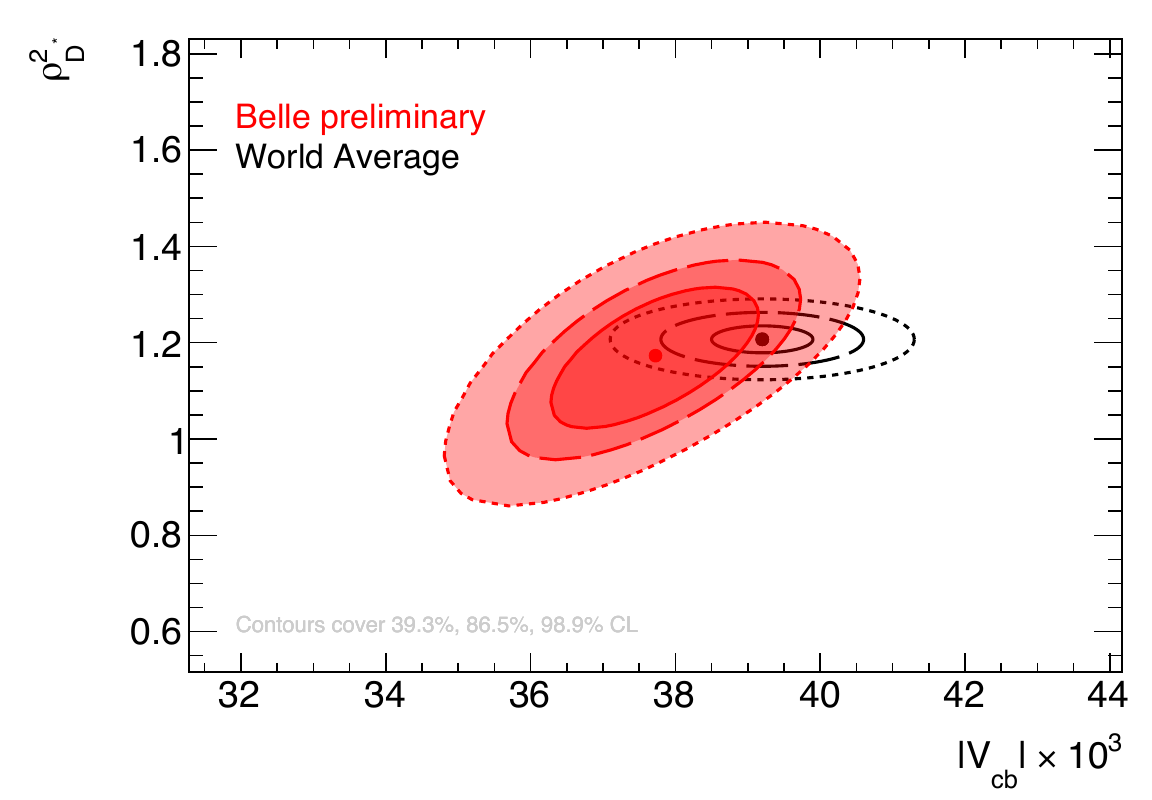} 
\includegraphics[width=0.48\textwidth,page=6,trim={0cm 0cm 0.5cm 0cm},clip]{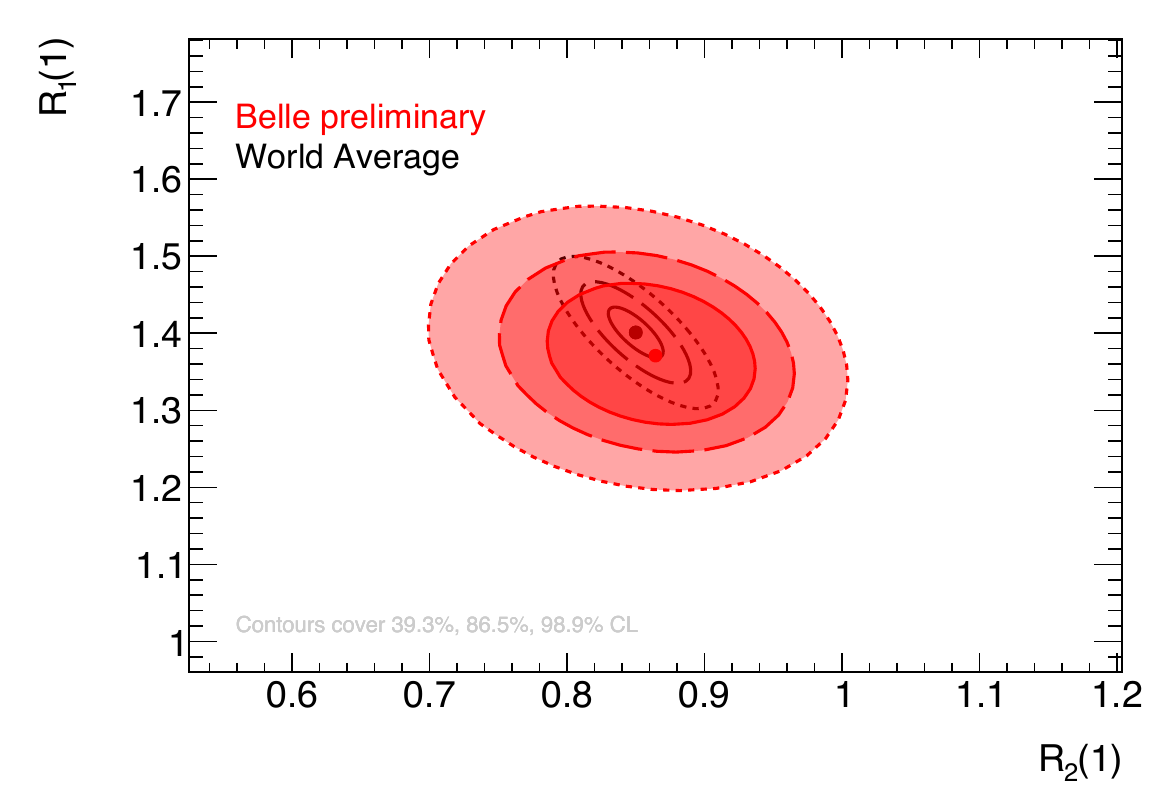} \\
\caption{ 
The best fit values for $\left| V_{cb} \right|$:$\rho_{D^*}^2$ and $R_1(1):R_2(1)$ with the corresponding $\Delta \chi^2 + 1$,  $\Delta \chi^2 + 2$, and $\Delta \chi^2 + 4$ contours are shown for the fit of the unfolded decay rates in red. Black shows the current world average from Ref.~\cite{hfag}.
}
\label{fig:FitRes2unfolded}
\end{figure}

\begin{figure}
\includegraphics[width=0.48\textwidth,page=1,trim={0cm 0cm 0.5cm 0cm},clip]{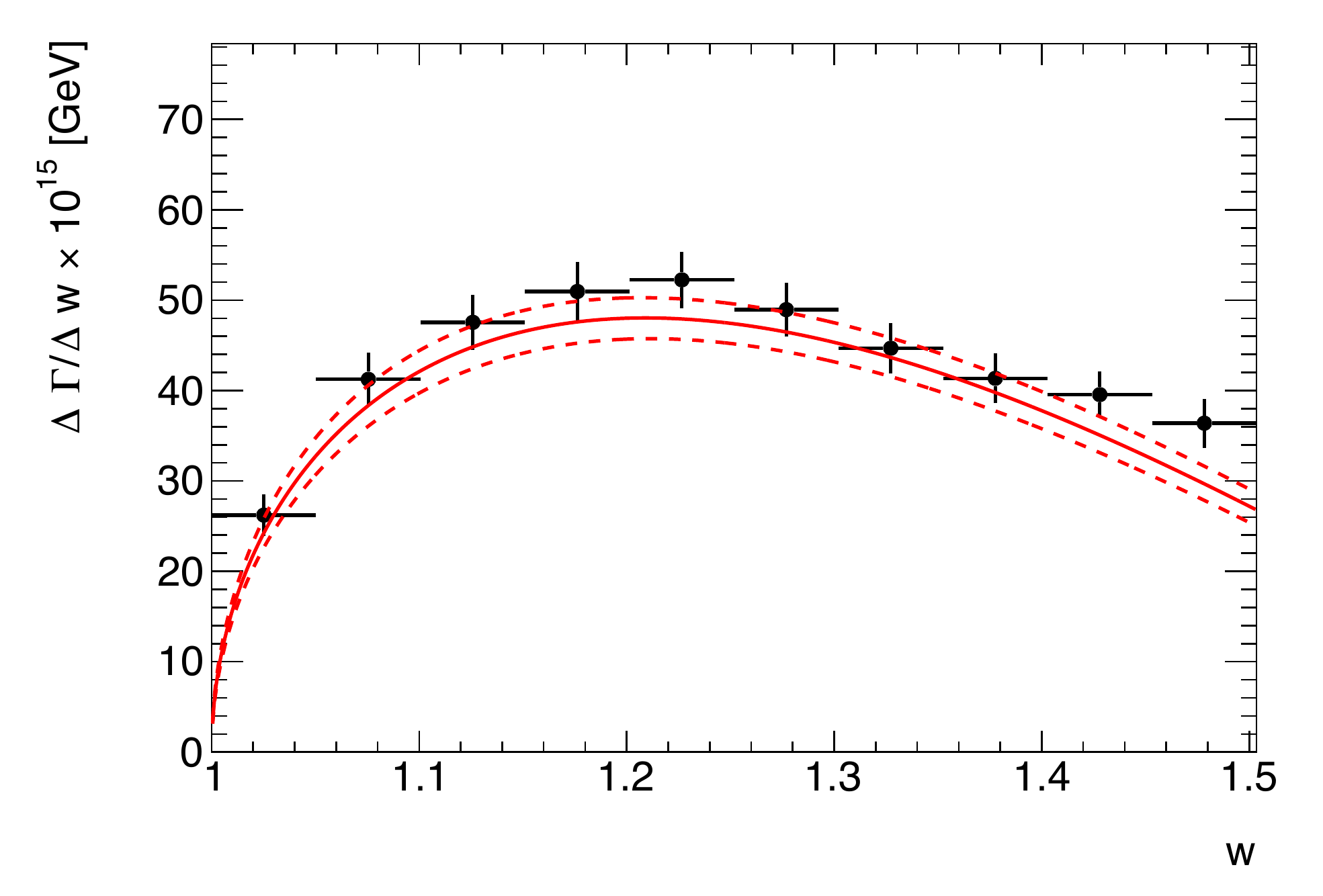} 
\includegraphics[width=0.48\textwidth,page=2,trim={0cm 0cm 0.5cm 0cm},clip]{./figures/B0_fitres_unfolded.pdf} \\
\includegraphics[width=0.48\textwidth,page=3,trim={0cm 0cm 0.5cm 0cm},clip]{./figures/B0_fitres_unfolded.pdf} 
\includegraphics[width=0.48\textwidth,page=4,trim={0cm 0cm 0.5cm 0cm},clip]{./figures/B0_fitres_unfolded.pdf} \\
\caption{ 
 The best fit values (solid red lines) and the corresponding $\Delta \chi^2 + 1$ errors (dashed lines) of the unfolded decay rates are shown. 
}
\label{fig:FitResunfolded}
\end{figure}

The unfolded fit result is summarized in Table~\ref{tab:fitsummaryunfolded} and Figs.~\ref{fig:FitRes2unfolded} and ~\ref{fig:FitResunfolded}. The fit has a $\chi^2 = 34.2$ with 40-4 degrees of freedom, corresponding to a fit probability of 47\%. The correlation matrix of the parameters  $\{ \left| V_{cb} \right|, \rho_{D^*}^2, R_1(1), R_2(1) \}$ is determined to be
\begin{align}
 C = & \left( \begin{matrix}
    1 & 0.63 & -0.04 & -0.37 \\
    0.63 & 1 & 0.20 & -0.83 \\
    -0.04 & 0.20 & 1 & -0.21 \\    
    -0.37 & -0.83 & -0.21 & 1 \\     
 \end{matrix} \right) \, .
\end{align}

\begin{table}[h]
\vspace{2ex}
\begin{tabular}{l|cc}
\hline\hline
Parameter & folded result & unfolded result   \\ \hline 
$\left| V_{cb} \right| \times 10^3$ & $37.4 \pm 1.3$ & $38.2 \pm 1.5$ \\
$\rho_{D^*}^2$& $1.04 \pm 0.13$ & $1.17 \pm 0.15$ \\
$R_1(1)$& $1.38 \pm 0.07$ & $1.39 \pm 0.09$ \\
$R_2(1)$& $0.86 \pm 0.10$ & $0.91 \pm 0.08$ \\
\hline\hline
\end{tabular}
\caption{ The best-fit values from fitting the unfolded spectra is shown and compared to the fit result using the less model dependent folding method.  }
\label{tab:fitsummaryunfolded}
\end{table}

\end{appendix}

\end{document}